\documentclass[12pt,preprint]{aastex}

\slugcomment{}

\shorttitle{Near-Infrared Observations of Globular Clusters in Early-Type 
Galaxies}
\shortauthors{Hempel et al.}

\begin{document}

\title{Near-Infrared Observations of Globular Clusters in NGC~4472,
NGC~4594, NGC~3585 and NGC~5813 and Implications for their Ages and
Metallicities} \author{M.~Hempel}\affil{Department of Astronomy,
University of Florida, Gainesville, FL 32611}
\email{hempel@astro.ufl.edu} \author{S.~Zepf and A.~Kundu}
\affil{Department of Physics and Astronomy,~Michigan State University,
East Lansing, MI 48824} \email{zepf,akundu@pa.msu.edu}
\author{D.~Geisler} \affil{Departamento de F\'isica,~Universidad de
Concepci\'on, Casilla 160-C, Concepci\'on, Chile }
\email{dgeisler@astro-udec.cl} \author{T.J.~Maccarone} \affil{School
of Physics and Astronomy, University of Southampton, Southampton,
Hampshire, SO17 1BJ, UK} \email{tjm@astro.soton.ac.uk}

\begin{abstract}

We present near-infrared photometry of the globular cluster systems of
the early-type galaxies NGC~4472, NGC~4594, NGC~3585, and NGC~5813. We
combine these near-infrared data, obtained with PANIC at the Magellan
Baade 6.5m telescope, with archival optical HST and FORS/VLT data, and
use the optical to near-infrared colors to constrain the ages and
metallicities of the globular clusters around the target galaxies.
For NGC~4472 we have the most extensive near-infrared and
optical photometric dataset. These colors show that the NGC~4472
globular cluster system has a broad metallicity distribution and that
the clusters are predominantly old (i.e. ages of about 10 Gyr or
more).  This result agrees well with earlier spectroscopic work on
NGC~4472, and is evidence that the combination of optical to
near-infrared colors can identify predominantly old systems and
distinguish these from systems with a substantial intermediate age
component. Based on the smaller sample of combined optical and
near-infrared data NGC~4594 and NGC~3585 appear to have
predominantly old globular cluster systems, while that of NGC~5813 may
have a more significant age spread. We also match our sample of
globular clusters with near-infrared and optical photometry to Chandra
X-ray source detections in these galaxies, and discuss how the
probability that a globular cluster hosts a low-mass X-ray binary
depends on metallicity and age.

\end{abstract}


\section{Introduction}
\label{introduction}

One of the primary goals in current extragalactic astronomy is to
determine the formation history of galaxies. The study of globular
cluster systems is a valuable tool for addressing galaxy formation and
evolution. One reason for this is that globular clusters are
simple stellar populations, with all of the stars in the cluster
sharing the same metallicity and age. Hence, determining these values
for an extragalactic globular cluster is more straightforward than
attempting to do this using the complicated mix of populations in the
integrated light of a galaxy. Moreover, because each cluster has a
distinct age and metallicity that can be determined from observations,
the distribution of ages and metallicities of a globular cluster
system can be derived.

One of the most promising ways to determine the ages and metallicities
of extragalactic star clusters is through combining optical and
near-infrared colors (e.g. Goudfrooij et al. 2001; Puzia et al. 2001;
Hempel et al. 2003). This technique overcomes the well-known
degeneracy between age and metallicity in optical colors alone
(Worthey 1994), taking advantage of the different sensitivities to age
and metallicity of both colors. Specifically, near-infrared colors
depend primarily on the giant branch and are therefore much more
sensitive to metallicity than age, while optical colors also probe
stars near the main sequence turnoff, and therefore have a significant
sensitivity to age as well as metallicity. The comparison between the
observed colors and those given by Simple Stellar Population models
(SSPs), e.g. by Bruzual \& Charlot 2003, allows the determination of
relative ages for any distinct sub-population. For example,
$(V-I)$~$vs.$~$(V-K)$ color distributions can be used to distinguish
globular cluster populations if their ages differ by several Gyr
(Puzia et al. 2002; Hempel \& Kissler-Patig 2004). The optical plus
near-infrared technique has been used successfully in the past few
years in studies of globular cluster systems. One of the notable
results from these studies are the detection of intermediate age
globular cluster populations in NGC~1316 (Goudfrooij et al. 2001),
NGC~4365 (Puzia et al. 2002; Hempel et al. 2003; Kundu et al. 2005),
and IC~4051(Hempel et al. 2005). Several other galaxies studied show
no such sign of intermediate age populations, such as NGC~3115 (Puzia
et al. 2002) and M87 (Kissler-Patig, Brodie and Minniti 2002) or only
small younger fractions, as found in NGC~1399 (Kundu et al. 2005;
Hempel et al. 2006). The detection of intermediate age populations
in the GCSs of some elliptical galaxies has shown that an exclusively
early formation of these galaxies in a single burst of star formation
followed by passive evolution can not account for all elliptical
galaxies. However, significantly larger galaxy samples are required to
answer questions such as the fraction of early type galaxies in the
local universe with such intermediate age globular cluster populations
and how the presence or absence of any significant age structure is
correlated with various parameters such as the environment of the
galaxy. The primary goal of the work presented here is to help build a
sample of local early-type galaxies with deep near-infrared and
optical photometry from which the age structure in their globular
cluster systems can be constrained.\\ The possibility of determining
the ages and metallicities of GCs around elliptical galaxies is also
interesting for the study of low-mass X-ray binaries (LMXBs) in these
galaxies. LMXBs are overabundant in GCs compared to the field
population and indicate the importance of dynamical formation of LMXBs
in the dense GC environment (e.g. Clark 1975; Katz 1975a; Katz
1975b). For the LMXBs in GCs, the information on the age and/or
metallicity opens up the unique opportunity to directly determine
these parameters for the host system, and thus assess their role of
these parameters in the formation and evolution of LMXBs. Therefore,
as in previous studies (Kundu, Maccarone, and Zepf 2002; Kundu et
al. 2003; Maccarone, Kundu, and Zepf 2003) we combine Chandra X-Ray
observations with the results of our optical/ near-infrared photometry
to determine the properties of the GC and LMXB matches.

The setup of this paper is as follows. Section \ref{obs} describes the
galaxy sample in which we study the GCs, the new near-infrared
observations presented in this paper, and their combination with
existing optical data to produce optical to near-infrared colors for
GCs in the target galaxies. The age results for the various globular
cluster systems are presented in section \ref{result}. Section
\ref{lmxbs} combines the results on the globular clusters with the
LMXBs detected in the target galaxies. A summary of our major results
is given in section \ref{discussion}.\\

\section{Observations}
\label{obs}
\subsection{Galaxy sample}
\label{sample}

The work presented here is part of a continuing study, investigating
the globular cluster systems of early-type galaxies, using globular
clusters as probes of major star formation events. The primary goal is
to use near-infrared and optical photometry to detect/identify any
sub-populations in the globular cluster systems, defined by their age
and metallicity. Here we present the data and results for four
additional early-type galaxies, differing to some extent in their
luminosity (-22.75$\leq$~M~$_{V}$$\leq$-21.79) and significantly in
local galaxy density (0.12$\leq$~$\rho$~$\leq$3.31
Mpc$^{-3}$). General information on our target galaxies- NGC~4594,
NGC~4472 and NGC~3585 and NGC~5813, is given in Table \ref{galaxies}.

\placetable{galaxies}

\subsection{Observational data and data reduction}

\subsubsection{Near-infrared data}
\label{nir}

The near-infrared K$_{s}$-band data were obtained with the
{\it{PANIC}} instrument (Persson's Auxiliary Nasmyth Infrared Camera)
during the nights of March 31, April 01 and April 02 in 2005. The
PANIC instrument (Martini et al. 2004) is mounted at the the 6.5 m
Magellan (Walter Baade) Telescope at the Las Campanas Observatory,
Chile. It is equipped with a Rockwell HgCdTe 1024x1024 Jugate
detector, with a pixel scale of 0.125''/pixel, yielding a 2\farcm0
x~2\farcm0 field of view. The observing strategy for the four galaxies
was the following: we observed a single central field with
5x(2x15~sec)~on source+ 5x(2x15~sec)~sky+ 5x(2x15~sec)~on source
exposures.  The sky exposures were taken at a position 5\farcm0 east
of the central field. For both on-source and sky observations we
applied a 5 point dither sequence with a 1\farcs0 dither. The total
on-source exposure times were: 2700 sec (NGC~4594), 4770 sec
(NGC~4472), 3690 sec (NGC~3585), and 5070 sec (NGC~5813)
respectively. Using the PANIC IRAF
\footnote{IRAF is distributed by the National Optical Astronomy
Observatories, which are operated by the Association of Universities
for Research in Astronomy, Inc., under cooperative agreement with the
National Science Foundation.} package (Martini \& Perrson 2004) the
exposures within one loop (2 readouts, 15 sec each) were co-added
({\it loopsum}), corrected for non-linearity ({\it lincor}),
flat-fielded ({\it flatten}) and sky-subtracted ({\it imarith}). The
individual exposures were aligned and shifted with respect to a
selected single exposure. The alignment of the exposures was checked
by comparing the new pixel coordinates of 7 (NGC~4594, NGC~4472) and
10 (NGC~3585, NGC~5813) objects on each individual exposure. The final
shifting and trimming of the exposures was done using the IRAF task
{\it{imalign}} and the total exposures created with {\it{imcombine}}
as the mean of the individual exposures.\\

The photometric conditions were monitored during the nights by
observing {\it{Persson}} standard stars from the {\it{Faint Standard
Star catalog}} (Perrson et al. 1998; Harwarden et al. 2001). The first
night as well as parts of the following two nights were found to be
photometric, and photometric calibration of data obtained under
non-photometric conditions was performed using shorter exposures of
the target galaxies taken under photometric conditions. Aperture
photometry on both the standard stars and the science exposures was
carried out using the Source Extractor program (Bertin \& Arnouts
1996). The following calibration relation has been derived for the
photometric calibration of the short/ photometric exposures:

\begin{eqnarray}
K=k_{inst}+ZP-0.08*\chi
\label{calibeq}
\end{eqnarray}

Hereby k$_{inst}$ denotes the instrumental magnitude, ZP the
photometric zero point and $\chi$ the effective airmass (Stetson
1988). According to Frogel (1998) we adopted an atmospheric extinction
coefficient of 0.08. The different calibration constants, as well as
the applied Galactic foreground reddening corrections (Schlegel,
Finkenbeiner, and Davies 1998) are given for each galaxy in Table
\ref{calibrate}. The mean photometric error for the K-band data,
e.g. in NGC~4472 is $\sigma (K)$=0.087 mag, as given by
SExtractor. Given the high accuracy of the optical photometry the
mean error in the visible band is $\sigma(V)$ = 0.02 mag, and
the mean error in $(V-K)$ is $\sigma(V-K)$= 0.09 mag.

\placetable{calibrate}

To calibrate the total exposures, the photometric offset between the
calibrated short and the total exposures, called {\it{offset}} in
Table \ref{calibrate}, was derived for each data set, as well as the
aperture correction between a 5 pixel (diameter) and an infinite
aperture. Using the IRAF task {\it{median}} we created a smoothed
image of the diffuse galaxy light which was subtracted from the total
exposure. The final photometry was eventually carried out on the
galaxy subtracted image using Source Extractor. Although the
calculation of the total exposure time for each target was meant to
ensure that equally deep data were obtained, the data quality still
varies. In our set of galaxies NGC~4472 is the class winner, with the
best, i.e. deepest and largest, near-infrared data set. The total
number of GCs in this galaxy is 37, after applying the error cut of
0.15 mag for both colors. The data set for NGC~4594 is of similar
quality, but a FOV mismatch between optical and near-infrared
observations results in a modest number of GCs (26 including selection
by error cut). In the NGC~3585 GC sample 26 objects are finally
included in the study. Data were also taken for NGC~5813, a bright,
large galaxy in the Virgo- Libra cloud (Huchra \& Geller
1982). However, the combined optical/ near infrared data set (error
selected) contains only 22 objects.\\

\subsection{Optical photometry}
\label{opt}

The globular cluster systems of all four galaxies have been
well-studied in the optical bands (HST/WFPC2), and we use the results
published by Kundu \& Whitmore (1998) for NGC~4594 and NGC~4472, and
Puzia et al. (2004) for NGC~3585 and NGC~5813, respectively. Table
\ref{optical} gives some details on the observations; we refer the
reader to the original publications for further details.

\placetable{optical}

To match the optical sources with our near-infrared (K-band) data, we
used the IRAF task {\it tfinder}.  The RA and DEC coordinates of the
globular clusters as given by HST/ACS, HST/WFPC2 or VLT/FORS2 were
transformed into pixel coordinates with respect to the {\it PANIC}
near-infrared observations. Finally both source lists, optical and
near-infrared, were matched allowing a maximum offset of 3 {\it PANIC}
pixels (0.375 arcsec).

\section {Results}
\label{result}

\begin{figure}[!ht]
\centering
\includegraphics[scale=0.5]{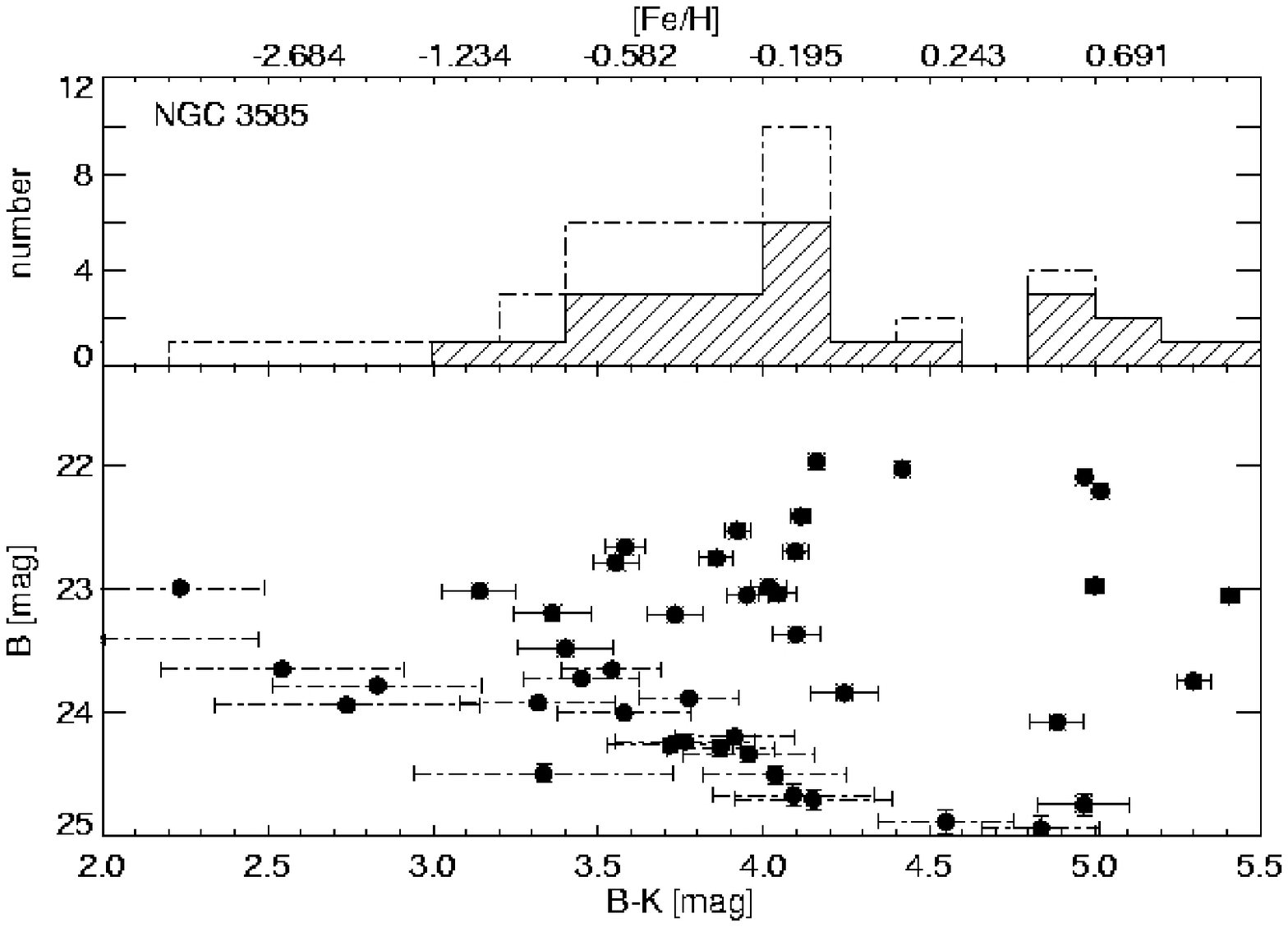}
\includegraphics[scale=0.5]{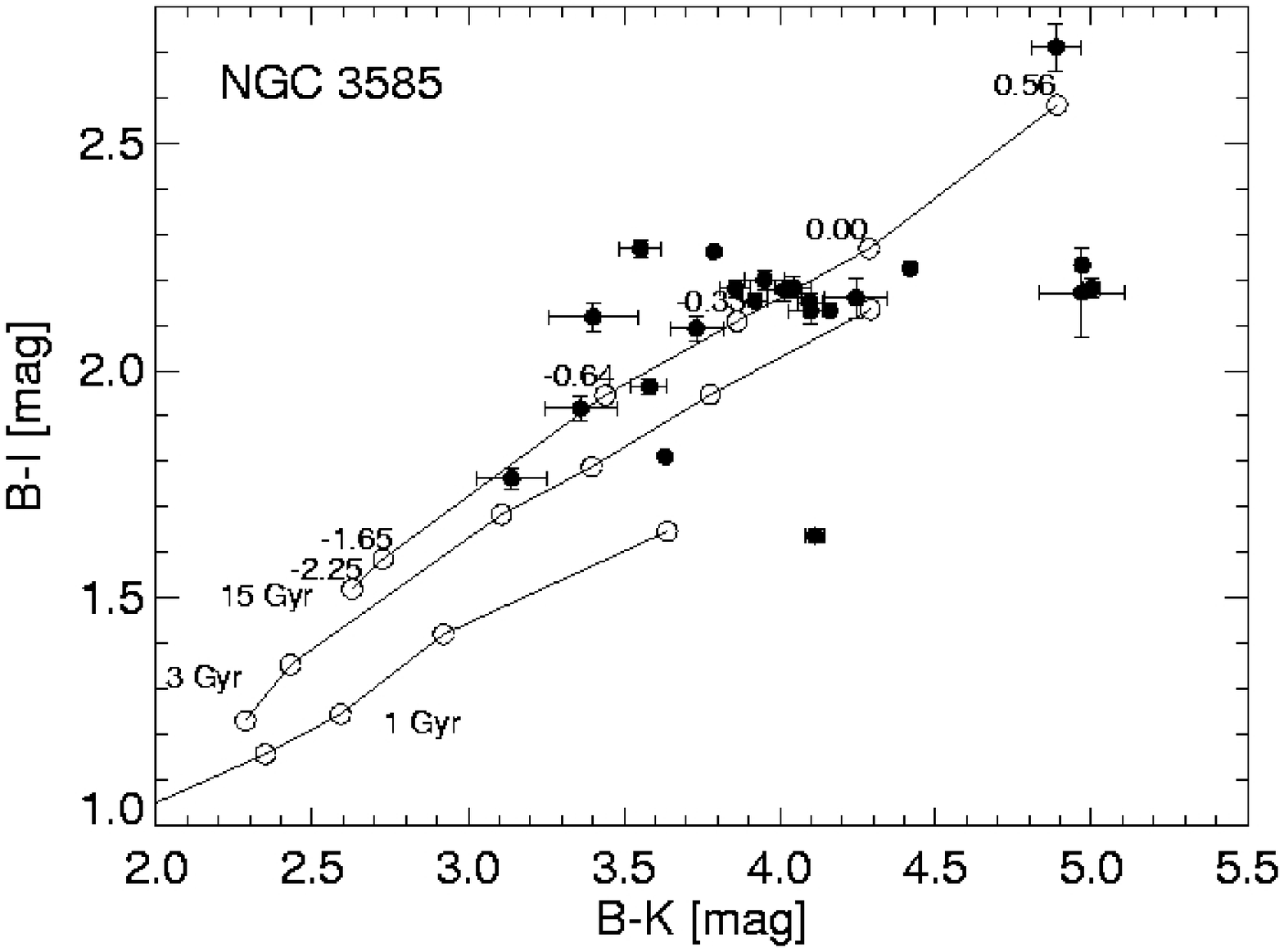}
\caption{Color magnitude and color-color diagram for
NGC~3585 GCs. Although the optical data include the B-band instead of the V-band,
the selection criteria remain the same (photometric errors for both
colors $\leq$0.15~mag). No strong evidence fo a significant population of intermediate age clusters is seen. }
\label{n3585}
\end{figure}

\begin{figure}[!ht]
\centering
\includegraphics[scale=0.5]{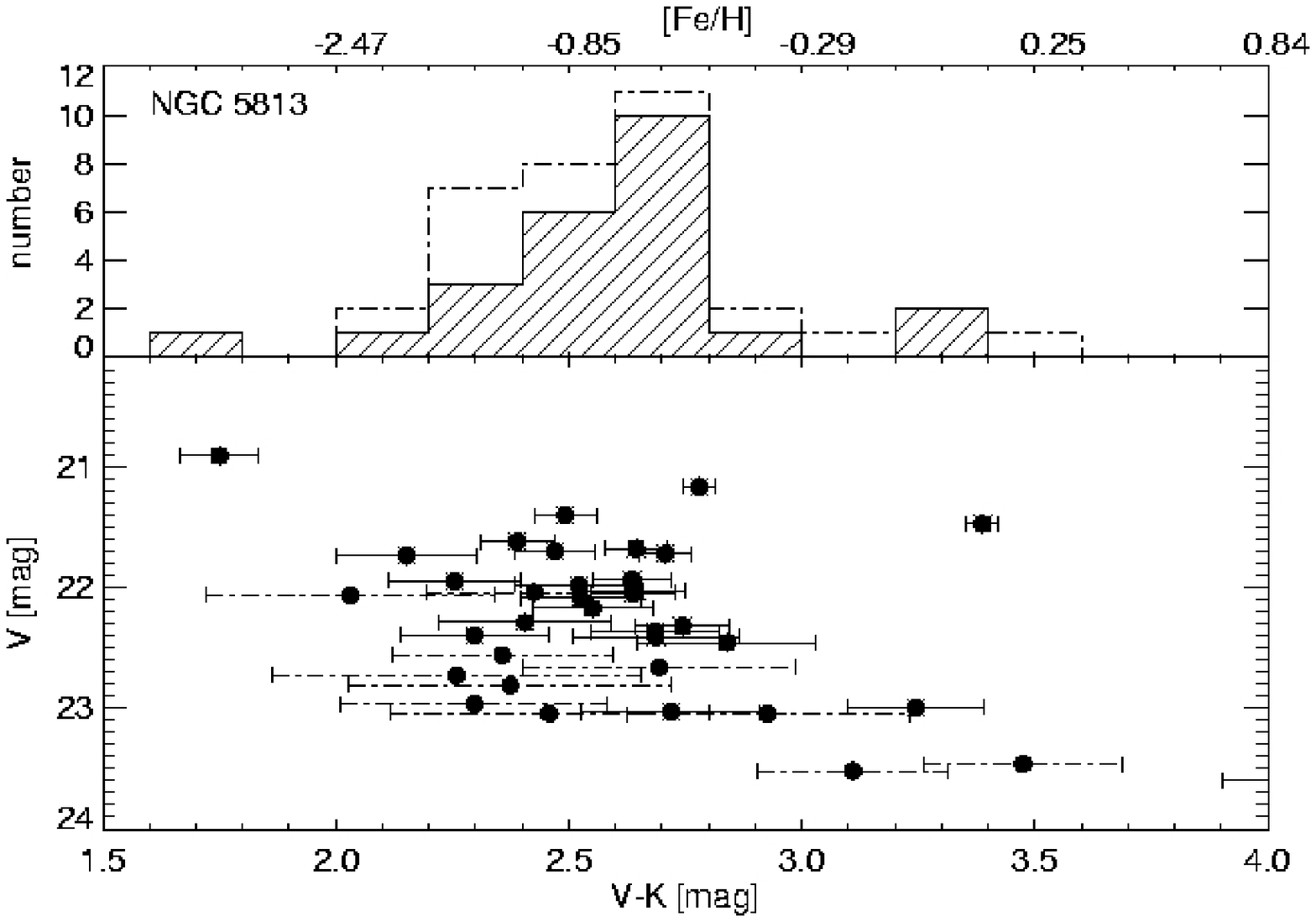}
\includegraphics[scale=0.5]{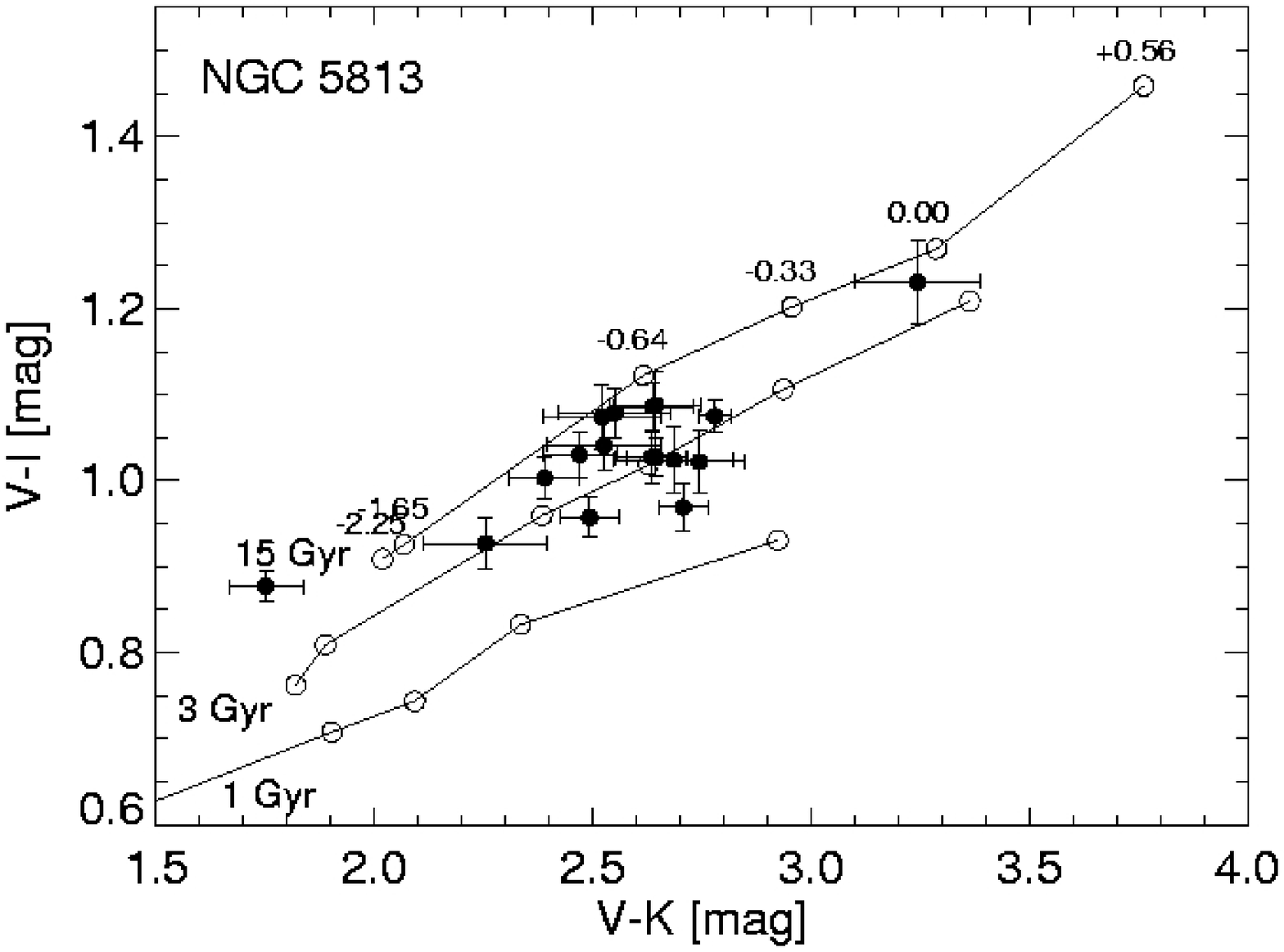}
\caption{Color magnitude and color-color diagram for NGC~5813 GCs. Given
the small number of objects, most of which are found within the blue
$(V-K)$ population, we tentatively suggest a second intermediate age
population of globular clusters. Of course, this needs to be confirmed
by deeper dataset. However, we note that the observed sample, small as
it is, lacks old, metal-rich clusters. The selection criteria as
well as the symbol settings correspond to the ones used for the other
target galaxies.}
\label{n5813}
\end{figure}

To set constraints on the globular cluster ages and metallicities, we
compare the $(V-I)$ $vs.$ $(V-K)$ two-color diagram to the predictions
of Single Stellar Population models. The color-magnitude diagrams,
color distributions and color-color diagrams of the four galaxies are
shown in Figures \ref{n4472}, \ref{n4594}, \ref{n3585} and
\ref{n5813}.  In this study we apply the models by Bruzual \& Charlot
(2003). In Hempel \& Kissler-Patig (2004) and Hempel et al. (2005) we
have shown that the main result regarding the presence or absence of
intermediate aged clusters is relatively model independent. In the
following we discuss the results for each of the galaxies
individually, starting with the galaxy with the largest dataset,
NGC~4472, followed by NGC~4594, NGC~3585, and NGC~5813, in order of
decreasing size of our globular cluster sample for each galaxy.
Symbols and selection criteria are set equivalently for all four
targets. In the CMDs globular clusters with photometric errors
$\Delta(V-I)$ and $\Delta(V-K)$ $\leq$0.15~mag are represented by the
solid points and solid line error bars. In the color distributions
these objects are represented by the solid line histograms. Objects
which have been rejected by the error cut are marked by open points
and dashed error bars in the CMDs and as open histograms in the
corresponding color distributions. As for the color-color diagrams:
solid lines represent the Bruzual \& Charlot SSP isochrones (Bruzual
\& Charlot 2003) for a 1,3, and 15 Gyr old population. For a given age
the metallicity increases from -2.25 to +0.56 with the $(V-K)$ color
index (open circles).

As briefly described in Section \ref{introduction}, it is possible to
disentangle age and metallicity effects in $(V-I)$,$(V-K)$ color-color
plots because the near-infrared light is primarily sensitive to
metallicity while the optical colors are sensitive to both age and
metallicity.  In general, predominantly old globular cluster systems
will follow the line of constant old age, with GCs of various
metallicities falling along the line from low metallicity on the
bottom left to high metallicity on the top right. In contrast, systems
with a significant intermediate age population will have substantial
numbers of GCs spread out in the age direction, along the isochrones
of younger age.  A key point is that calibration issues in either the
models or the data will shift a single age population relative to the
lines of constant metallicity but will not produce a dispersion in the
age direction. Greater spreads can be produced by large random errors,
but these should generally scatter about a specific age line for a
single age population. Moreover, the comparison between the dispersion
in GC populations in the age direction can be done on a relative basis
between galaxies studied in the same way, giving confidence that any
age spread in a given galaxy is real if such a distribution of
GC colors is not observed in similar observations of other GC systems.

\subsection{NGC~4472}
\label{res4472}

\begin{figure}
\centering
\includegraphics[scale=0.5]{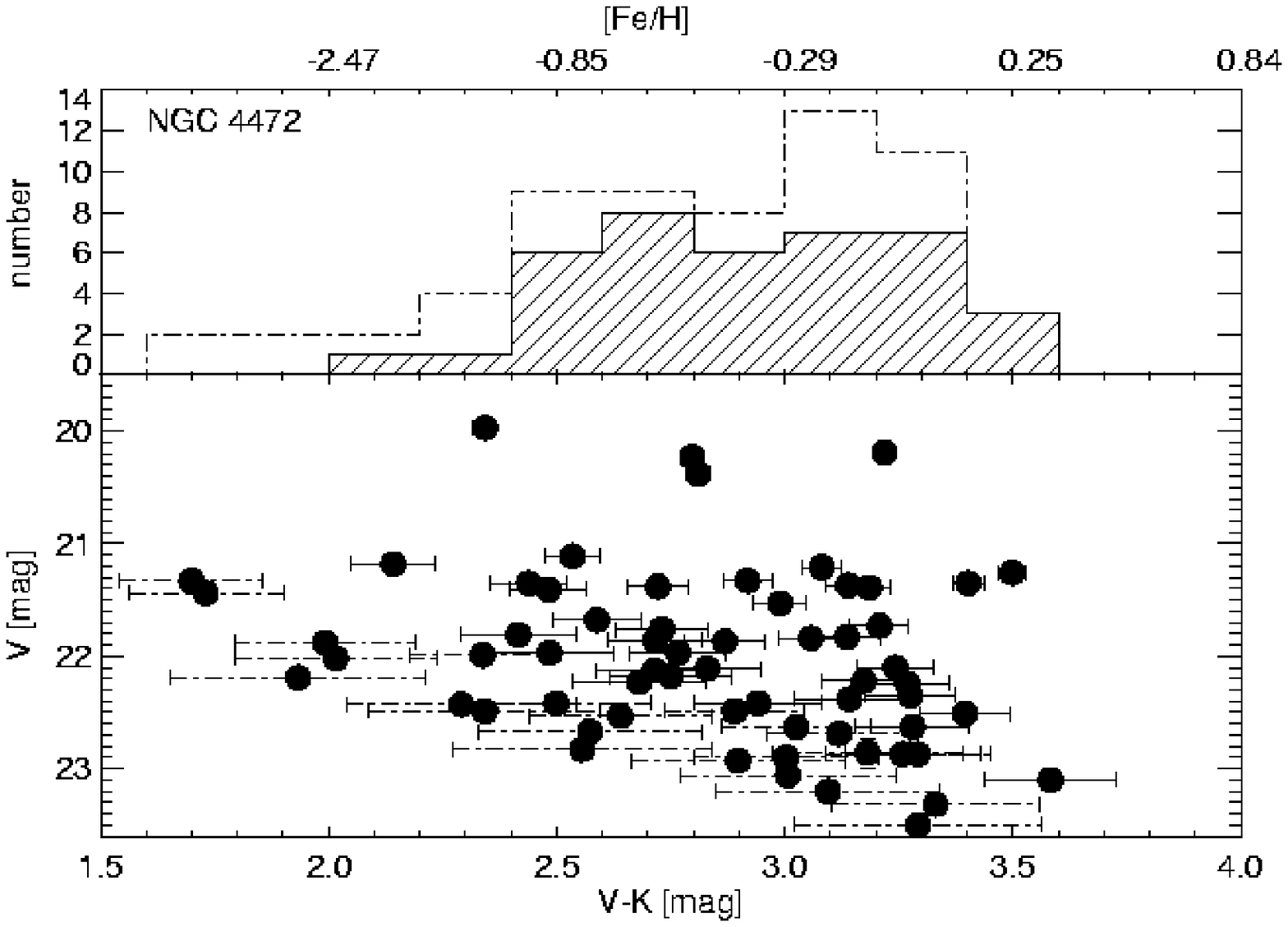}
\includegraphics[scale=0.5]{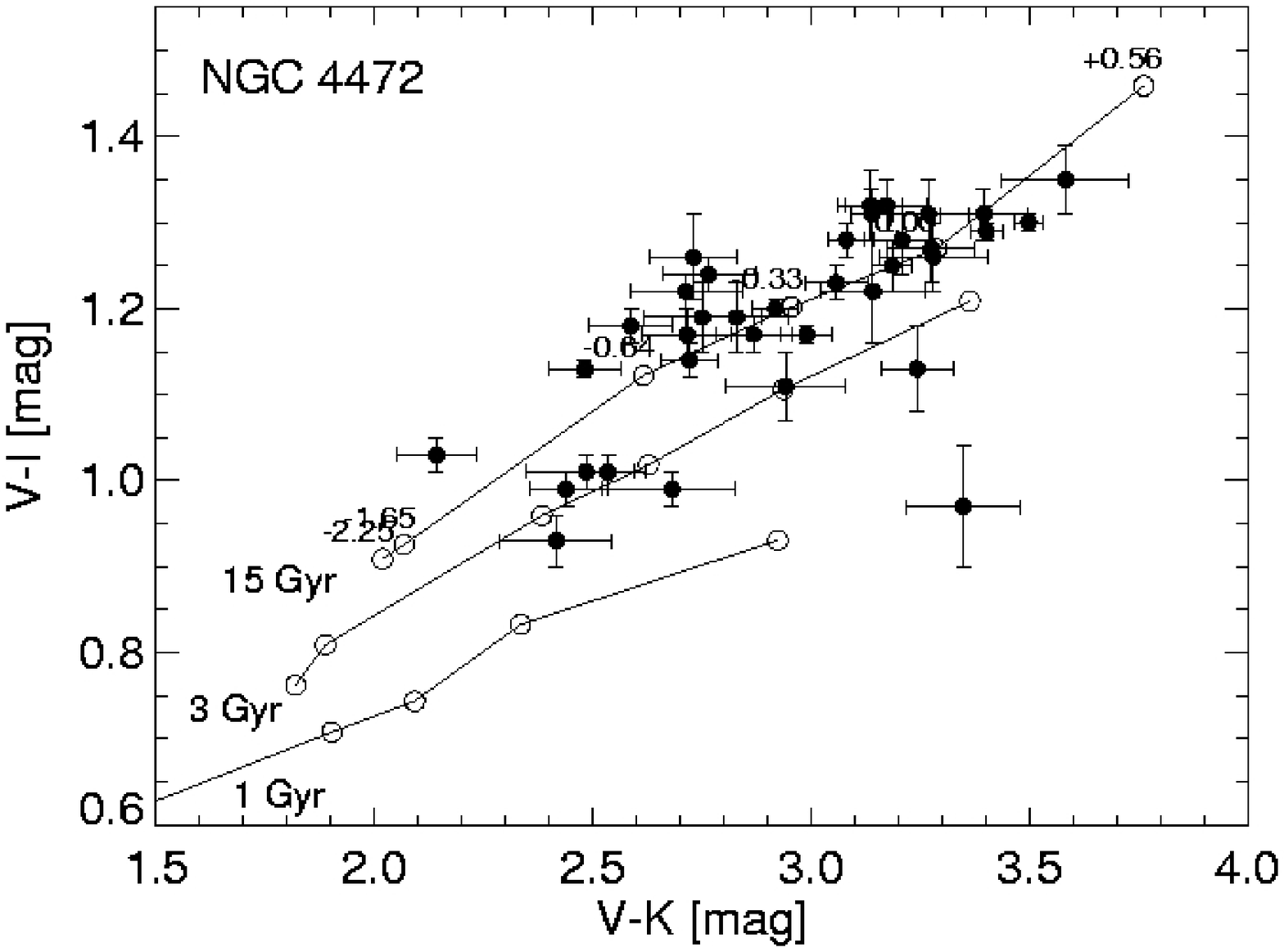}
\caption{Color histogram, color magnitude diagram and color-color diagram for NGC~4472 GCs. The
objects marked by solid circles in the color-magnitude diagram as well
as the solid line histogram refer to the error-selected objects
($\Delta$$(V-I)$, $\Delta$$(V-K)$$\leq$ 0.15~mag) whereas the dashed
lines refer to objects rejected by the error selection. The solid
lines in the color-color diagram represent the 1, 3 and 15 Gyr
isochrone following the Bruzual \& Charlot (2003) SSP model. The
metallicity is given for the 15 Gyr isochrone only. The metallicity scale on the upper panel is derived by interpolating the metallicities for a 15 Gyr old population folllowing the Bruzual \& Charlot isochrones (2003).}
\label{n4472}
\end{figure}

\begin{figure}[!ht]
\centering
\includegraphics[scale=0.5]{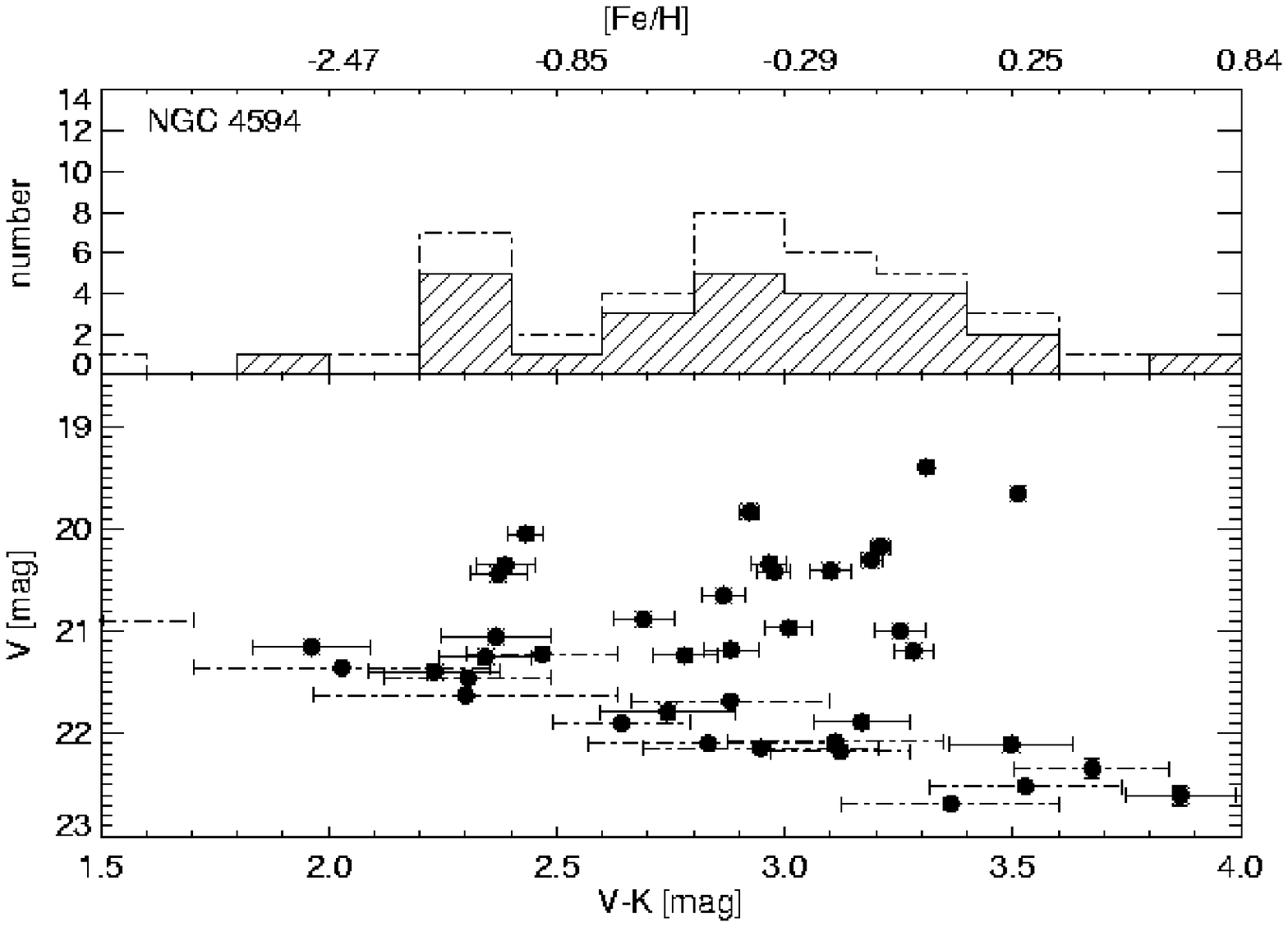}
\includegraphics[scale=0.5]{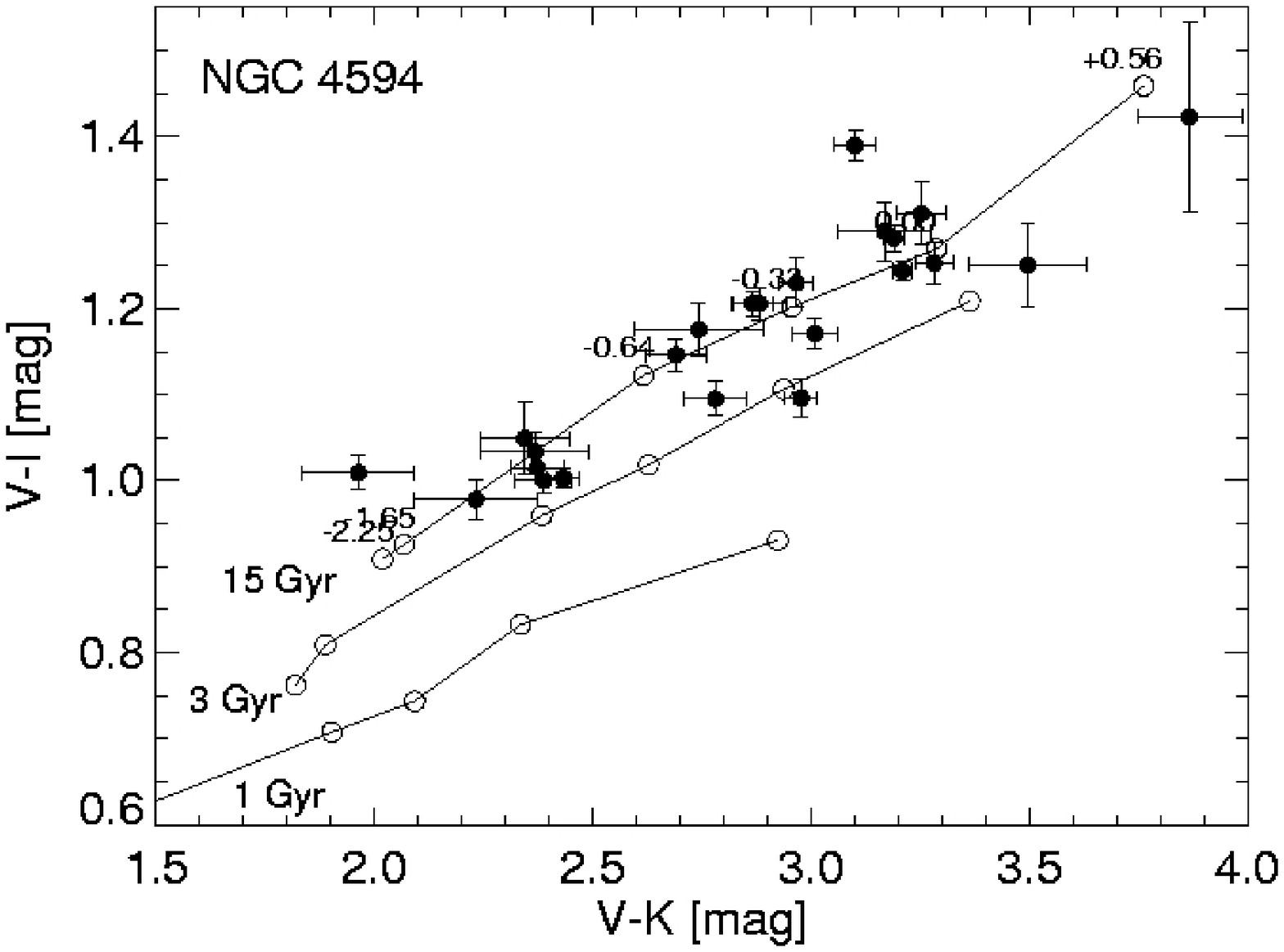}
\caption{Color magnitude and color-color diagram for NGC~4594 GCs. The
settings for symbols and lines as well as the selection criteria
correspond to Figure \ref{n4472}. Based on the color distribution we
find the GCs in this sample to be mostly old, with a wide spread of
metallicity.}
\label{n4594}
\end{figure}

NGC~4472 (M49) is a very interesting target for our study, as it is
the most luminous galaxy in the Virgo cluster, but does not reside at
the gravitational center of the host cluster. The globular cluster
system has been observed both photometrically (e.g. Geisler, Lee, and
Kim 1996; Rhode \& Zepf 2001; Kundu \& Whitmore 2001a) and
spectroscopically (e.g. Beasley et al. 2000; Cohen, Blakeslee, and
C{\^o}t{\'e} 2003). Comparing the optical color distribution (Kundu \&
Whitmore 2001a) with its counterpart in the near-infrared (see also
table \ref{photo4472}) we find the $(V-K)$ color distribution (see
Figure \ref{n4472}, upper panel) to be bimodal, similar to the optical
color distribution seen in earlier studies in Kundu \& Whitmore
(2001a). However, using the Hartigan \& Hartigan test for
unimodality (DIP-test) results in a dip-value of 0.0625. For our
sample of 37 objects (no color limits, only error cuts are applied) we
obtain a probability of 70 $\%$ that the $(V-K)$ color distribution is
not unimodal. In comparison, if we fit the color distribution with
single or double Gaussian we obtain $\chi$$^2$-values of 0.85 and
1.27, respectively. Both methods are hampered by the small number of
objects in the sample. Comparing the GC colors, shown in the lower
panel of Figure \ref{n4472}, with the SSP models lead to the
conclusion that the NGC~4472 GCS is built by a predominantly old
($>$10 Gyr) stellar population. This result is in good agreement with
the spectroscopic age estimates by Beasley et al. (2000) and Cohen,
Blakeslee, and C{\^o}t{\'e} (2003) who derive the mean GC age to be
older than 10 Gyr. In order to set also some constraints on the
metallicity distribution, the color-magnitude diagrams (Figures
\ref{n4472}, \ref{n4594}, \ref{n3585}, \ref{n5813}) show the
metallicity values (top x-axis), corresponding to the $(V-K)$ color
for a 15 Gyr isochrone, following the Bruzual \& Charlot SSP
models (2003).

\begin{figure}[!ht]
\centering
\includegraphics[scale=0.5]{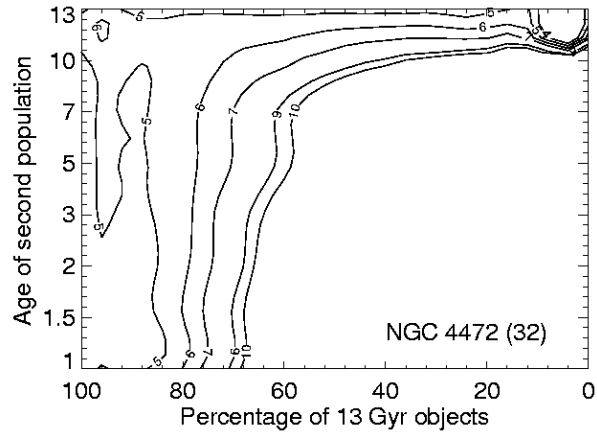}
\caption{Result of the $\chi$$^2$- test comparing the cumulative age
distribution derived for the red population of GCs (in $(V-K)$-color)
with simulated GCSs. After applying color and error cuts (Hempel \&
Kissler-Patig 2004) the sample contains 32 objects. As shown clearly
by the $\chi$$^2$ contours the age structure within the sample is best
fitted by an old ($\ge$10 Gyr) stellar population.}
\label{contourn4472}
\end{figure}

The size and quality of the NGC~4472 GC sample allows us to extend our
analysis to include the cumulative age distribution (Hempel et
al. 2003; Hempel \& Kissler-Patig 2004). The comparison with age
distributions derived for simulated GCSs with a given age structure
(age, size of age sub-populations) via a reduced $\chi$$^2$-test
allows us to find the best fitting model and hence to set some
constraints on the relative age and size of possible cluster
sub-populations. Hereby we assume that only red, supposedly metal-rich
GCs might form an intermediate age population. For bluer objects the
color difference between a 13 Gyr and 2 Gyr isochrone, in
comparison with the photometric errors, does not allow us to discriminate
between those ages.  In our analysis we therefore select only
objects with a $(V-K)$ color$\ge$2.6. Using the Bruzual \& Charlot
(2003) SSP models this corresponds to a metallicity of [Fe/H]$\approx$-0.7,
assuming an old population (13-15 Gyr). Figure \ref{contourn4472} shows
the result of the $\chi$$^2$-test for the NGC~4472 GC sample. The
contours represent the various $\chi$$^2$ levels. As in previous work
we use 88 different models, combining a 13 Gyr old population with a
second, intermediate age population of 1,~1.5,~2,~3,~5,~7,~10,~13
Gyr). For each model we assume a mixture of both age populations,
varying between a 100 $\%$ old and a 100 $\%$ intermediate population
(size increments: 10$\%$). The most important ingredient for our
method of photometric age estimates in GCSs is the combination of
optical and near-infrared data so we can take advantage of the
much reduced age -metallicity degeneracy. However, we still have to
accept the drawback of an increased photometric uncertainty, mostly
due to the ground-based infrared observations. This is the main reason
why individual GC ages as well as absolute GC ages are still out of
reach. Although we set an upper limit for the photometric error (0.15
mag) we still have to take the latter into account. To do so we use
the catalog of photometric errors of our observations. After populating
the primary color (i.e. $(V-K)$) randomly and calculating the
secondary, age sensitive, color index $(V-I)$ based on the SSP model
predictions, we smear both colors with an up to 3$\sigma$ photometric
error. The cumulative age distribution for the simulated systems is
than based on the newly calculated colors. 

Applying the above described procedure to the NGC~4472 globular
cluster sample finds the best fitting model (lowest $\chi$$^2$ value)
to consist only of objects older than 10 Gyr (see Figure
\ref{contourn4472}, top right corner). The photometric scatter seen in
Figure \ref{n4472} (lower panel) leads to the result that an equally
good fit to the age distribution is obtained when mixing a 13 Gyr old
population with a small fraction ($\leq$20$\%$) of intermediate age
GCs (younger than 7 Gyr). The different $\chi$$^2$-values vary between
4.63 (minimum) for a 90:10$\%$- mixture of a 13 and 2 Gyr population
and 5.05 for a combination of 90$\%$ of 13 Gyr old clusters and 10$\%$
of a 10 Gyr old population. Under the assumption of only 10$\%$ young
objects we find that an age variation between 1 to 13 Gyr results in
very similar $\chi$$^2$ values, i.e. 4.79 (1 Gyr) and 4.68 (13
Gyr). Considering that different runs of simulations based on the same
age structure (e.g. purely 13 Gyr old population) cover a similar
$\chi$$^2$-interval (4.65 to 4.77) we find that a size resolution of
10$\%$ or better is not feassible for this sample. Nevertheless the
major outcome of this analysis is that NGC~4472 is by far dominated by
an old stellar population.

We note that good agreement between the photometric and spectroscopic
results, as is the case for NGC~4472, has also been found for NGC~1399
(Kundu et al. 2005; Hempel et al. 2006). This is of great importance
since it has been suggested that the optical/near-infrared method is
prone to misinterpreting photometric uncertainties as age spread,
hereby deriving an artificial second age population within the
globular cluster system (Brodie et al. 2005). However, our NGC~4472
results, as well as those for NGC~1399 (Kundu et al. 2005) demonstrate
that the near-ir to optical color technique is not only sensitive to
age spreads of several Gyr but also able to recover the age structure
of old stellar systems.

\subsection{NGC~4594}
\label{res4594}

NGC~4594 is a highly luminous field galaxy, with an absolute magnitude
of M$_{V}$=-22.4.  It has a bulge to disk ratio of 6.14
(Kent 1988), indicating its total luminosity is dominated
by its very bright bulge component, rather than its well-known edge-on
disk.  Given its high spheroid luminosity, field environment, and
proximity, NGC~4594 is a very interesting target for our study.

The globular cluster system of this galaxy has been observed
photometrically in the optical by Rhode \& Zepf (2004) and
spectroscopically by Larsen et al. (2002). The latter derive a GC age
around 10-15 Gyr. Using the combined optical and near-infrared colors,
shown in the lower panel of Figure \ref{n4594} (data are given in
Table \ref{photo4594}), we find the majority of the globular clusters
to be old ($\geq$10 Gyr), which is also in good agreement with the
spectroscopic results. Similar to the spectroscopic sample (14
confirmed GCs) the sample size in our study is not sufficient to rule
out completely the possibility of any intermediate age GCs in
NGC~4594, but any such population is clearly not a large fraction of
the total, if present at all.  Further tests of the age structure in
the NGC~4594 GC system using additional, wide-field near-infrared data
we have recently obtained are underway.


\subsection{NGC 3585}
\label{res3585}

NGC~3585 is another of the field galaxies in our sample and therefore
a useful extension of our galaxy sample when it comes to investigating
the influence of the galaxy environment on galaxy evolution. The local
galaxy density for NGC~3585 is the lowest in this sample (see Table
\ref{galaxies}). The color-color distribution of the NGC~3585 globular
cluster system, shown in Figure \ref{n3585} and Table \ref{photo3585},
does not show signs of a substantial intermediate age component. This
result is in agreement with previous work by Puzia et al. (2004, 2005)
using optical photometry and spectroscopic data.

\subsection{NGC~5813}
\label{res5813}

The globular cluster system as well as the integrated light of
NGC~5813 have been investigated extensively in the optical (e.g. Hopp,
Wagner, and Richtler 1995; Kundu \& Whitmore 2001a; de Zeeuw et
al. 2002; Howell 2005). The results, e.g. a dynamically de-coupled
core, and dust filaments in the inner center of the galaxy were early
on interpreted as reminiscent of a merger (Kormendy 1984), which since
has neither been confirmed (Howell 2005) nor securely ruled out (Hopp
1995). Hence NGC~5813 is an interesting target when searching for age
sub-populations in the GCS.  As shown in Figure \ref{n5813} the
globular cluster sample (18 objects) is dominated by a large fraction
of blue ($(V-K)$~$\leq$~2.6), globular clusters. Comparing the optical
and near-infrared colors (see Table \ref{photo5813}) of the GCs with
the SSP models leaves room for an intermediate age population although
the small number of objects, especially within the metal rich region
hampers the interpretation. Given the various galaxy properties and
our preliminary and not very conclusive results, NGC~5813 remains a
prime target for GC studies, which would require deeper near-infrared
data built on either additional long exposure and/or wider-field
ground-based data in good seeing or space-based observations
(HST/NICMOS). In our galaxy sample NGC~5813 represents galaxies in a
low density environment, similar to NGC~3585.


\section{Low Mass X-Ray Binaries in Globular Clusters}
\label{lmxbs}

Early-type galaxies are not only large concentrations of stellar mass
but also major sources of X-ray emission.  Chandra studies have
revealed that in many ordinary elliptical galaxies, much of this
emission comes from individual low-mass X-ray binaries (LMXBs)
(e.g. Trinchieri \& Fabbiano 1985; Sarazin, Irwin, and Bregman
2000). Further work has shown that many (about 40\%) of these LMXBs
are located in globular clusters (e.g. Angelini, Loewenstein, and
Mushotzky 2001; Kundu, Maccarone, and Zepf 2002; Kundu, Maccarone, and
Zepf 2006; Fabbiano 2006). The much higher number of X-ray sources per
unit stellar mass in globular clusters than in the field is strong
evidence that most or all of these systems are formed dynamically
through tidal captures, exchange interactions, or direct collisions of
compact objects with stars (e.g. Clark 1975; Katz 1975a,b; Fabian,
Pringle, and Rees 1975; Verbunt 1987).

Moreover, since less than one percent of the light of an elliptical
galaxy originates from star clusters, it follows that LMXBs are about
two orders of magnitude more likely to be found in GCs than in the
field. This strongly points to an important role for the dynamical
formation of LMXBs in globular clusters, as has also long been
realized from Galactic studies of LMXBs and globular clusters
(e.g. Bhattacharya 1995, and references therein). Because of the
strong connection between globular clusters and LMXBs, we can use our
knowledge about globular clusters to gain information about the
formation and evolution of the LMXBs.

Chandra data is available for three of the four galaxies in our
sample, NGC~4472, NGC~4594, and NGC~3585.  The galaxies were observed
with {\it{Chandra}} on June 12, 2000, May 31, 2001, and June 03, 2001,
respectively. The total integration time was 39.6 ksec, 20.0 ksec, and
35.7 ksec.

\begin{figure}[!ht]
\centering
\includegraphics[scale=0.3]{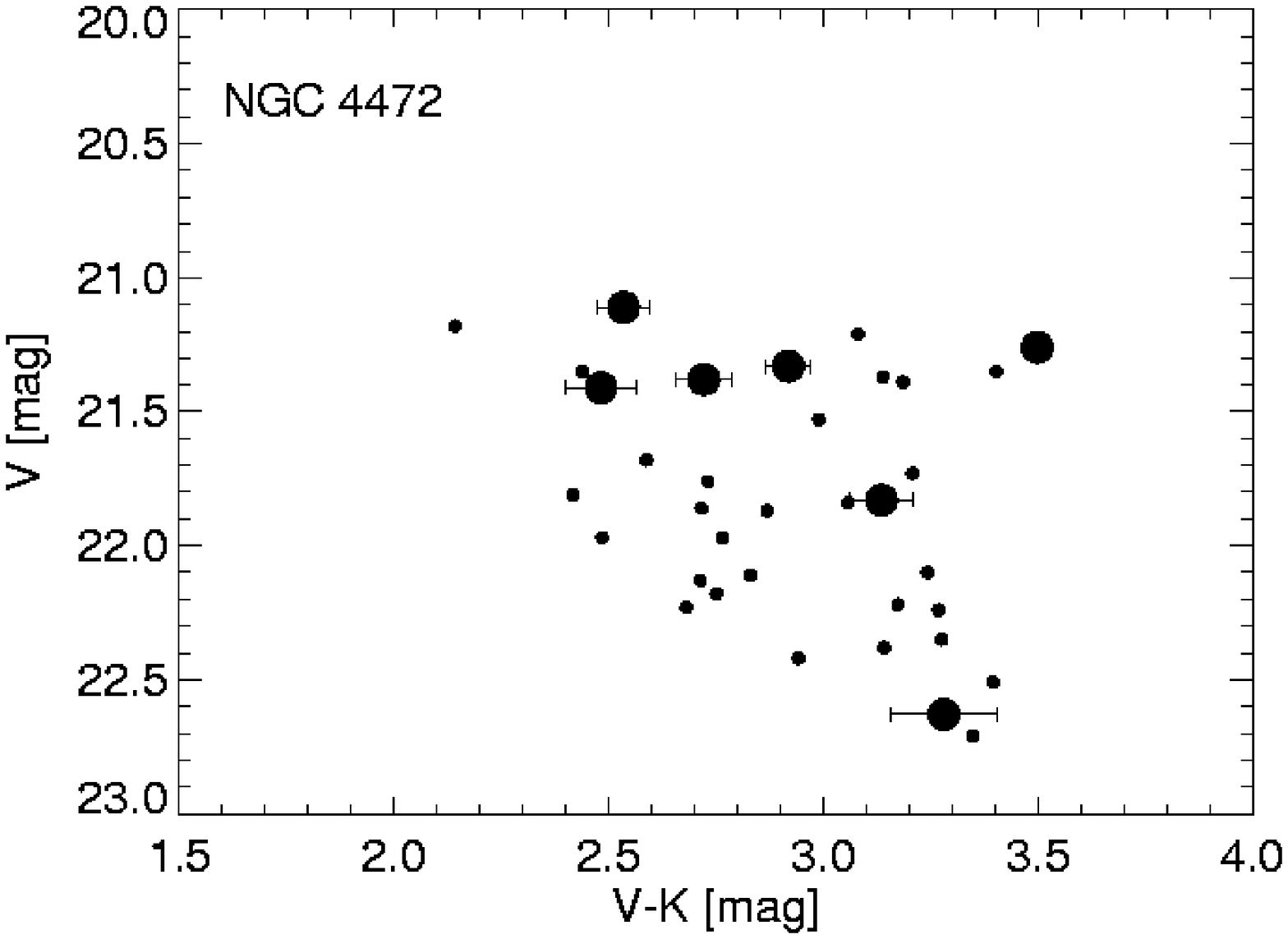}
\includegraphics[scale=0.3]{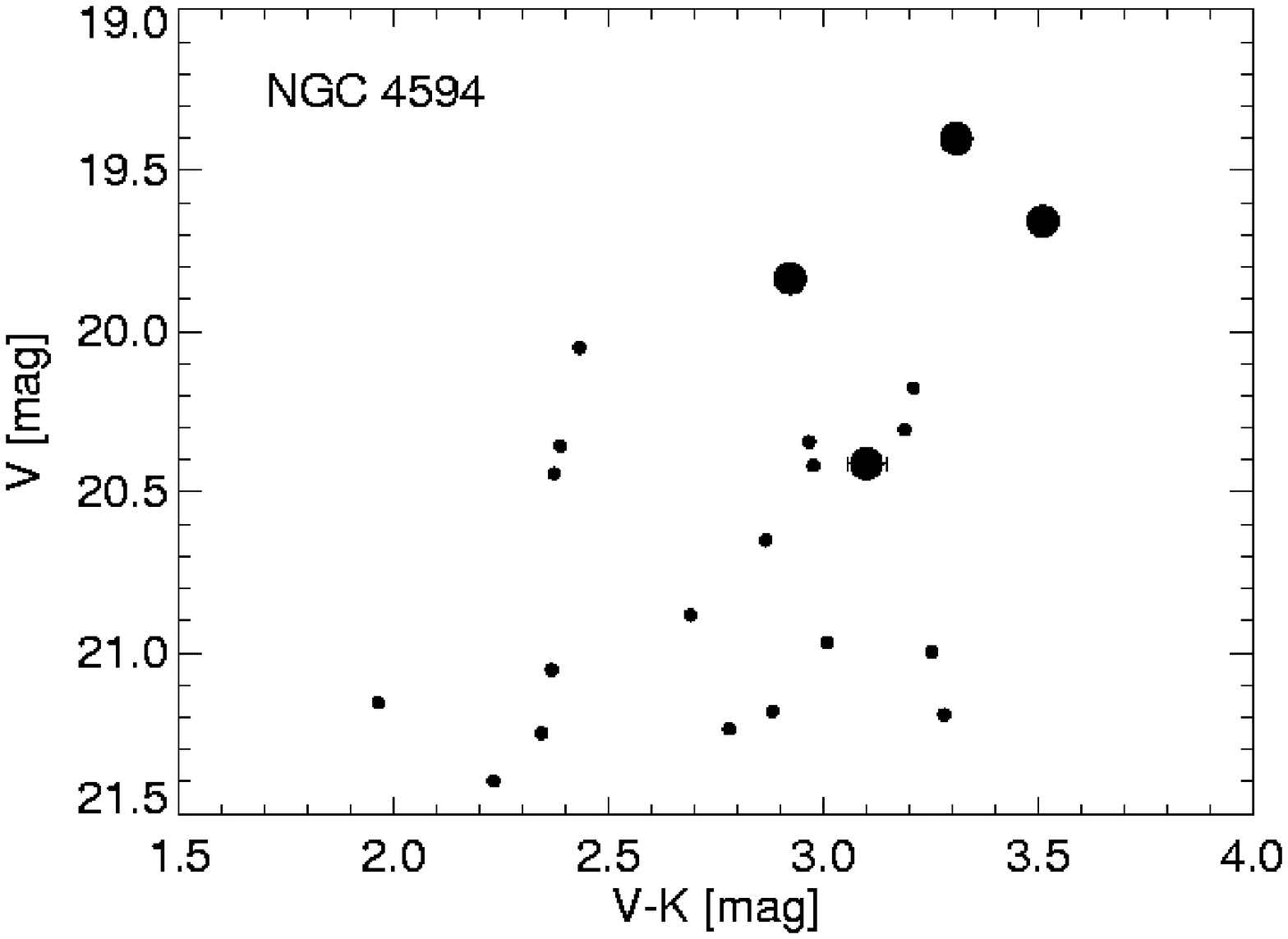}
\includegraphics[scale=0.3]{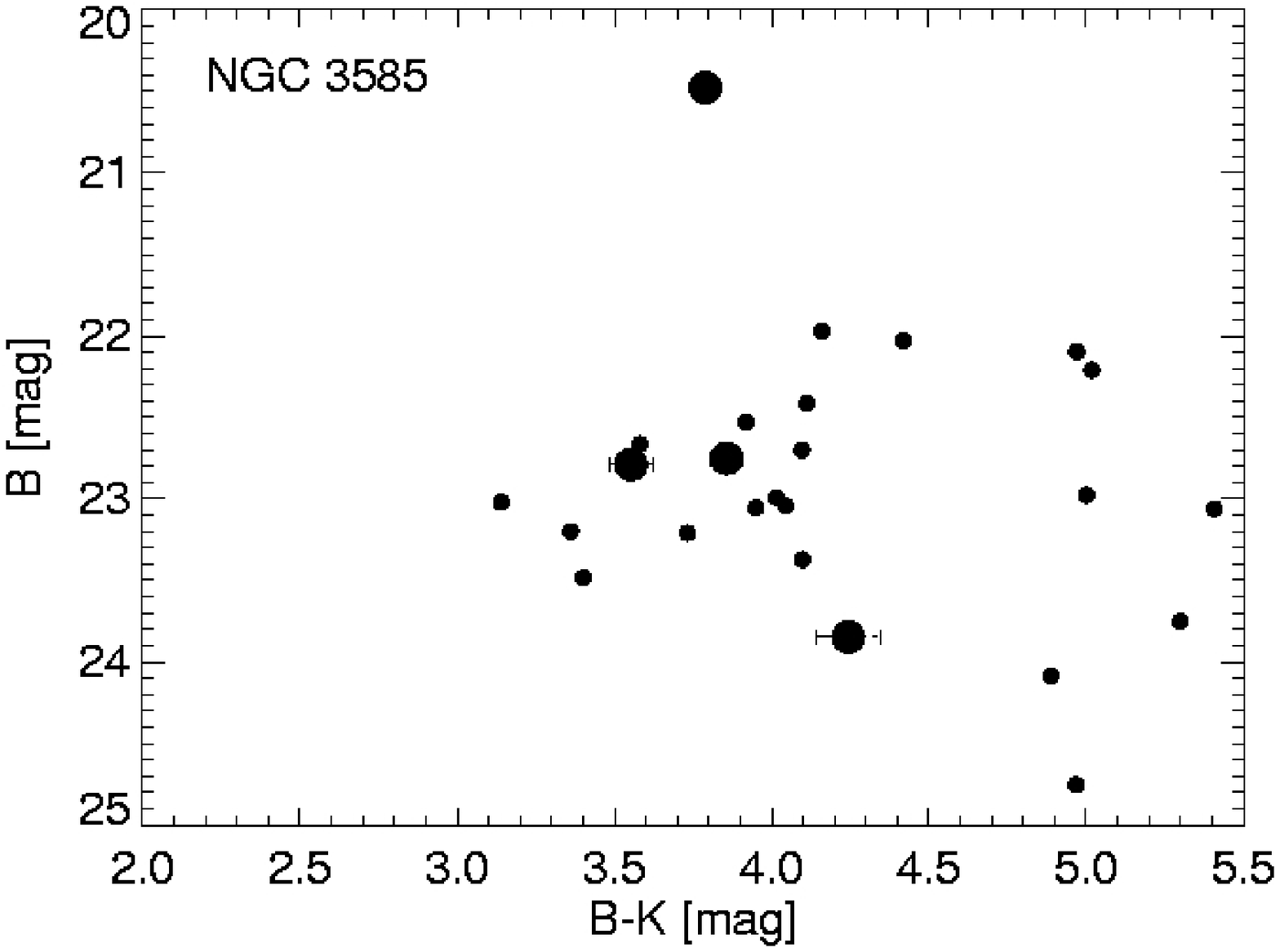}
\caption{Color-magnitude diagram for GCs
in NGC~4472 (left), NGC~4594 (center) and NGC~3585 (right). Only GCs
following the selection criterion (i.e. photometric errors for both
colors $\leq$0.15~mag) are plotted. The large symbols represent GCs
with a LMXB detection whereas small symbols are GCs without LMXB.}
\label{lmxbcmd}
\end{figure}

\begin{figure}[!ht]
\centering
\includegraphics[scale=0.3]{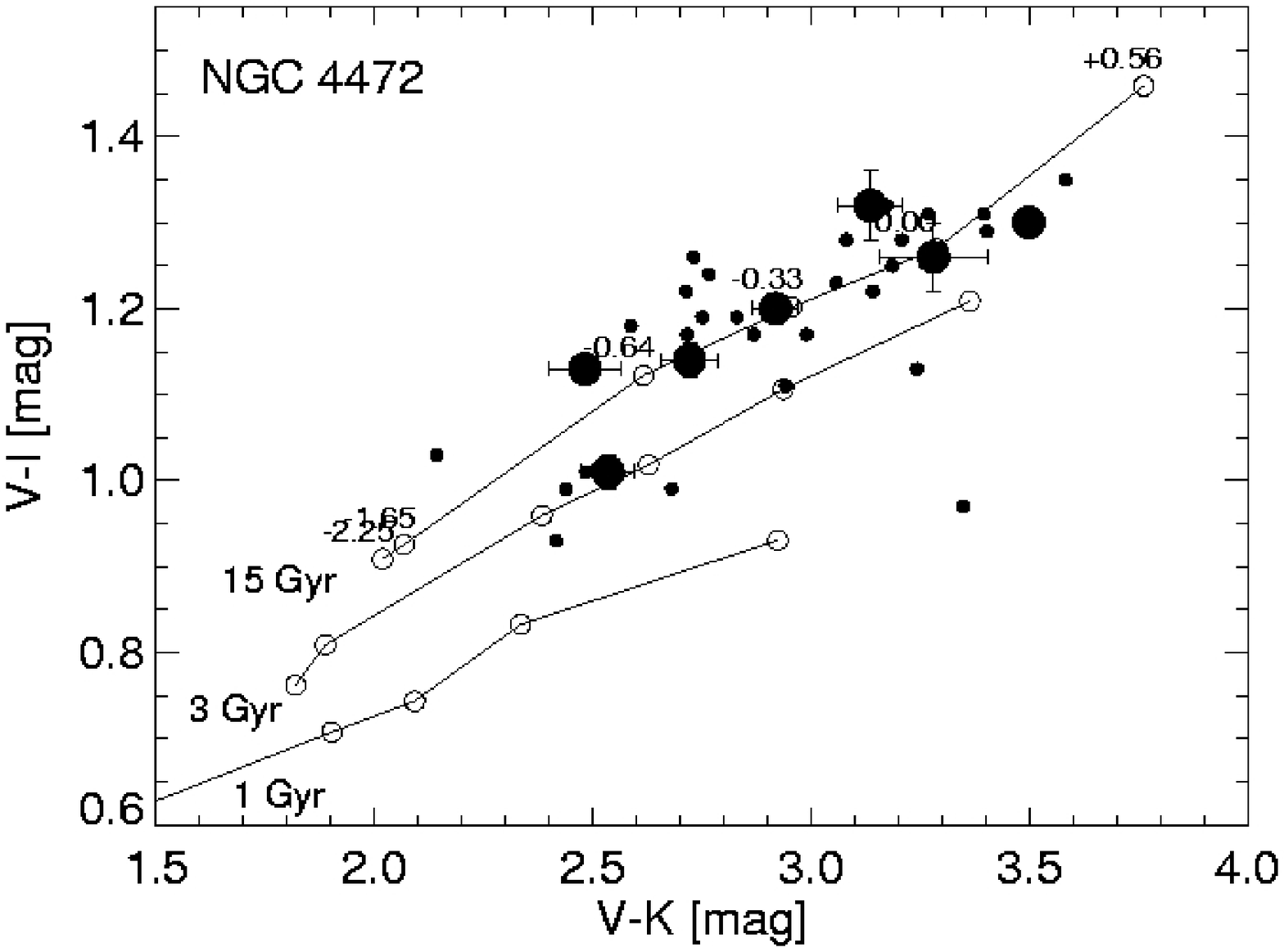}
\includegraphics[scale=0.3]{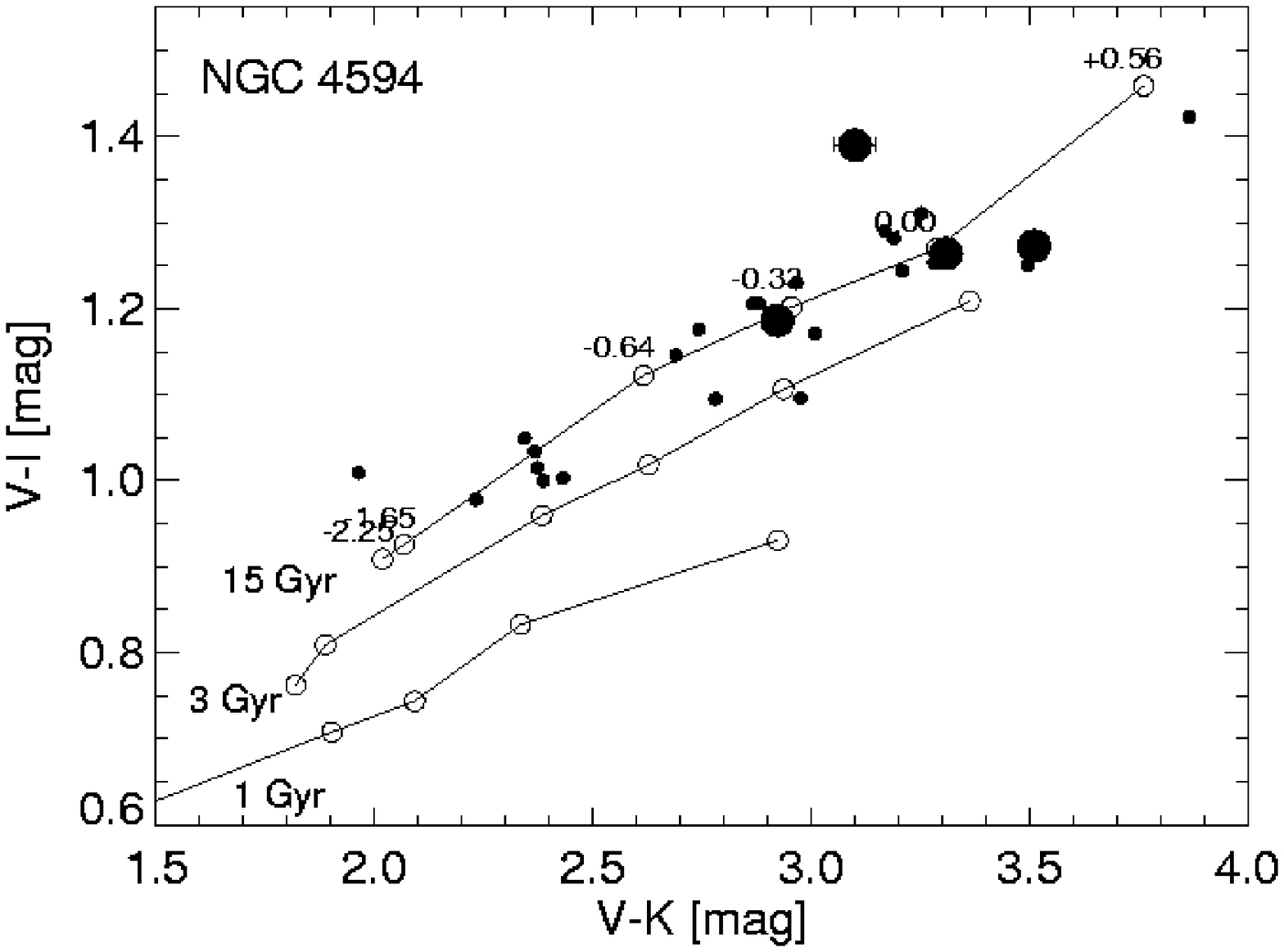}
\includegraphics[scale=0.3]{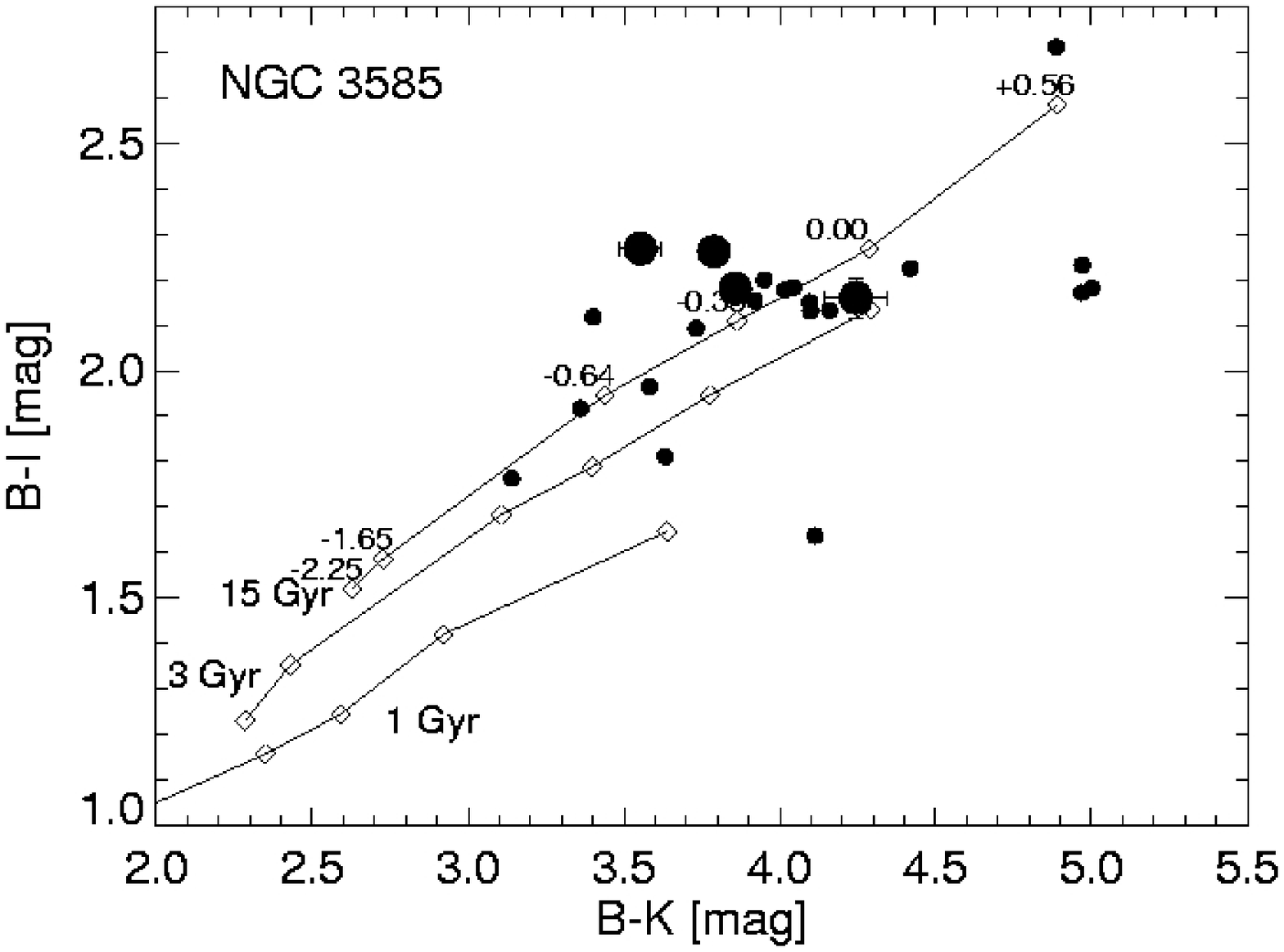}
\caption{Color-color diagram for GCs hosting LMXBs in NGC~4472 (left),
NGC~4594 (center) and NGC~3585 (right). GCs hosting an LMXB are marked
as large dots whereas small ones represent the GCs without LMXB. Only
objects obeying the selection criterion (i.e. photometric errors for
both colors $\leq$0.15~mag) are plotted. The SSP model isochrones are
by Bruzual \& Charlot (2003). The open circles mark the various
metallicities. GCs hosting LMXBs are predominatly old and metal-rich.}
\label{lmxbcol}
\end{figure}

To match the LMXBs with the GC catalog we use the list of objects
detected in the range of 0.5 to 8.0 keV. Standard data reduction
procedures were used for filtering the X-ray data for high background
intervals, making images, and extraction source lists. These
procedures are the same for all galaxies as they were for NGC~4472
(for details see Maccarone, Kundu, and Zepf 2003).  In Figures
\ref{lmxbcmd} and \ref{lmxbcol} LMXBs associated with globular
clusters are plotted (large symbols) together with the GCs without
LMXB detections (small symbols). Similar to the combination of the
optical and near-infrared data (see Section \ref{opt}) we used the
IRAF task {\it{tfinder}} to match the Chandra detections with the
globular cluster photometry. With both data sets centered on the
galaxies this study is somewhat hampered by the underlying galaxy
light. Including the relatively small size of the FOV (2'$\times$~2')
in the K-band observations we match 7, 4 and 3 (4) LMXBs in NGC~4472,
NGC~4594 and NGC~3585, respectively. We note that the brightest object
(B$<$20.5 mag) in NGC~3585 may not be a GC. Based on the apparent
B-band magnitude and the distance modulus given in Table
\ref{galaxies}, we derive that this cluster is about 4 times as
luminous as $\omega$CEN, the Milky Way's most luminous globular
cluster. Although it is by far the brightest object in terms of B-band
magnitude (20.47 mag) we find the corresponding X-ray source to be
the faintest of the four detections
(L$_{X}$$\sim$8.4*10$^{37}$erg/sec). Tentatively we would therefore
exclude the possibility of this object being an AGN and not a globular
cluster.\\



Based on the color-color plots (see Figure \ref{lmxbcol}), most of the
LMXBs in our sample are found in old and metal-rich globular clusters,
which is in agreement with previous studies
(e.g. Angelini et al. 2001; Kundu et al. 2002; Sarazin et al. 2003; Smits et al. 2006). With two
exceptions (in NGC~4594 and NGC~3585) LMXBs are found in bright GCs
(see Figure \ref{lmxbcmd}). Although our data allow a first comparison
between LMXB and non-LMXB globular clusters (see Figures
\ref{n4472},\ref{n4594} and \ref{n3585}), numerous questions about the
connection between GC properties and the formation efficiency of LMXBs
remain. For example, does the formation efficiency scale directly with
[Fe/H] or does the LMXB formation merely require a minimum
metallicity?

\section{Discussion and Summary}
\label{discussion}

We have combined our near-infrared with existing optical photometry of
the globular cluster systems of NGC~4472, NGC~4594, NGC~3585, and
NGC~5813 to investigate the ages and metallicities of the globular
cluster systems in these galaxies. One of the goals of this work is to
help build the sample of galaxies with useful constraints on the age
distribution of their globular clusters so that comparisons can be
made between the formation history of their host galaxies and other
galaxy properties such as mass and environment. The four galaxies
studied here help achieve this goal by spanning a range of galaxy
luminosities and environments from the Virgo cluster to lower density
field regions. 

The very high luminosity early-type galaxies in our sample, NGC~4472
and NGC~4594, both appear to have globular cluster systems that are
premarily old ($>$10~Gyr) with a broad metallicity distribution. This
old age is clearly established in the Virgo elliptical NGC~4472 and
also is very likely for the field early-type galaxy NGC~4594. The
other two galaxies in the sample, NGC~3585 and NGC~5813, are
elliptical galaxies with fairly typical luminosities found in the
field or group environments. The modest sizes of the globular cluster
samples with good optical to near-infrared colors for these two
galaxies mitigate against strong conclusions, but the current data
suggest NGC~3585 may be primarily old while NGC~5813 may have a
significant intermediate age component.

It may be notable that of the four galaxies, the one with some
evidence for a substantial intermediate age component, NGC~5813, is
 not in a cluster of galaxies and also not the most luminous galaxy in
our sample. This is very broadly consistent with the idea that on
average less luminous early-type galaxies in lower density
environments are more likely to have had more recent formation
activity (e.g. Kuntschner et al. 2002; Thomas et al. 2005, de Lucia et al. 2006, Sanchez-Blazques et al. 2006).

The existence of spectroscopic studies of some of the globular cluster
systems studied here allows for an independent test of the
near-infrared to optical photometric technique. For NGC~4472, the old
age we find for the globular clusters is in excellent agreement with
previous spectroscopic studies (Beasley et al. 2000, Cohen, Blakeslee,
and C{\^o}t{\'e} 2003). Similarly, while the numbers are not as large
in both the spectroscopic and photometric datasets, the old ages we
find for the NGC~4594 globular clusters also agree with earlier
spectroscopic work (Larsen et al. 2002). Although with very few
numbers, spectroscopic studies of NGC~3585 globular clusters (Puzia et
al. 2005) are also consistent with our results.  Moreover, our
detection of galaxies with mostly old ($\geq$10 Gyr) GCSs as well as
the discovery of other galaxies which have GCSs with intermediate age
sub-populations through the same technique (e.g. NGC~4365, Puzia et
al. 2002; Hempel \& Kissler-Patig 2004; Kundu et al. 2005) lend confidence to the supposition that the near-infrared to optical colors are identifying
real differences in the GC systems for which the only known
explanation is age.

A central long-term goal of this work is to compare the formation
histories of early-type galaxies indicated by the age structure in
their globular cluster systems to various theoretical models. An
obvious step to reaching this goal is to expand the sample size of
early-type galaxy GCSs with good near-infrared to optical photometry,
of which this work is part. There are also several other steps that
will be helpful in reaching this goal. One of these is obtaining
wider-field near-infrared and optical photometry of a subset of these
galaxies.  Nearly all of the current constraints are for the inner few
arc minutes corresponding to no more than several galactic effective
radii. While this is in part because that is where many of the GCs
are, it is also the case that GC systems extend to larger radii, and
wider-field near-infrared and optical photometry will be very useful
to assess whether the inner regions are fully representative of the
whole GC system.  Another important step for comparing GC systems of
different ages is to properly account for dynamical evolution.  This
will require combined progress in both theoretical modeling and
observational study of GC systems of a variety of ages. We also note
that photometric studies can be complemented by spectroscopic data,
obtained for selected globular clusters of each cluster
sub-population. This provides an independent estimate of the ages and
metallicities, and also potentially allows constraints on the duration
of the star formation episode using the $\alpha$-~enhancement in the
GCs.

For NGC~4472, NGC~4594, and NGC~3585, we also match our near-infrared
and optical GC photometry to Chandra observations of LMXBs in these
galaxies. This allows us to probe the connection between GCs and
LMXBs and test which properties of GCs are correlated with the
presence of LMXBs.  Utilizing near-infrared to optical colors for this
work has two advantages. One is the ability to separate the effects of
age and metallicity, thereby allowing age effects to be tested. The
second advantage is improved metallicity determinations, both because
any age effect in the colors can be accounted for and because of
improved precision in the metallicity due to the larger wavelength
baseline. We clearly confirm previous results that metal-rich GCs are
much more likely to host LMXBs than metal-poor GCs. Additional data
will be required to test whether this dependence on metallicity is
roughly proportional to metallicity, or whether most of the effect is
accounted for in the division between metal-poor and metal-rich
populations.

\begin{acknowledgements}
MH  and SEZ acknowledge support from NASA LTSA grant NAG5-11319, Chandra
grant AR5-6013X, and NSF award AST-0406891.
AK was supported by NASA LTSA grant NAG5-12975.  D.G. gratefully
acknowledges support from the Chilean {\sl Centro de Astrof\'\i sica}
FONDAP No. 15010003.  

\end{acknowledgements}

{}


\begin{table}
\centering
\caption[width=\textwidth]{General information on the host
  galaxies. Data were taken from: \\
(1) Tonry et al. (2001), \\
(2) de Vaucouleurs et al. (1991), \\ 
(3) Tully \& Fisher (1988), \\
(4) Pahre (1999),\\
(5) Schlegel, Finkenbeiner, and Davies (1998) \\}
\label{galaxies}
\begin{tabular}{l r r r r}
\hline 
\noalign{\smallskip} 
Property                & NGC~4594 & NGC~4472 & NGC~3585  & NGC~5813 \\
\hline
\noalign{\smallskip} 
$(m-M)$ (1)             & 29.95    & 31.06     & 31.51   & 32.54     \\
M$_{V}$ (2)             & -22.4    & -22.75    & -21.92  & -21.79    \\
$\rho$~[Mpc$^{-3}$] (3) &  0.32    &  3.31     &  0.12   &  0.88     \\ 
$(V-I)$$_0$ (1)         &  1.175   &  1.218    &  1.160  &  1.189    \\ 
D$_{25}$ (4)            &  8.40    & 11.40     &  4.50   &  3.70     \\ 
A$_{B}$ [mag] (5)       &  0.221   &  0.096    &  0.276  &  0.246    \\
\noalign{\smallskip} \hline
\end{tabular}
\end{table}

\begin{table}
\centering
\caption[width=\textwidth]{K-band calibration parameters for the
various galaxies. The values for galactic reddening were taken from
\citet[][]{schlegel98}.\\}
\label{calibrate}
\begin{tabular}{l r r r r}
\hline 
\noalign{\smallskip} 
                       & NGC~4594 & NGC~4472 & NGC~3585 & NGC~5813 \\
\hline
\noalign{\smallskip} 
ZP$_{phot}$ [mag]      & 24.398    & 24.300     & 24.415  & 24.411   \\
$\Delta$ZP$_{phot}$ [mag] &  0.014 &  0.043     &  0.043  &  0.030   \\ 
effective airmass~$\chi$&  1.233    &  1.338     &  1.267  &  1.294   \\ 
offset [mag]& 24.395$\pm0.009$ & 24.385$\pm0.045$ & 24.402$\pm0.022$& 24.407$\pm0.016$   \\
A$_{K}$ [mag] (1)       &  0.019    &  0.008     &  0.023  &  0.021   \\
\noalign{\smallskip} \hline
\end{tabular}
\end{table}

\begin{table}
\centering
\caption[width=\textwidth]{Exposure times for the optical photometry
on NGC~4472, NGC~4594, NGC~3585 and NGC~5813\\ (1) VLT/FORS2 \\ (2)
HST/WFPC2; F555W \\ (3) HST/WFPC2; F658W \\ (4) HST/WFPC2; F814W }
\label{optical}
\begin{tabular}{l r r r r}
\hline 
\noalign{\smallskip} 
                      & NGC~4472  & NGC~4594   & NGC~3585    & NGC~5813\\
\hline
\noalign{\smallskip}      
B                & --        & --         & 800 sec (1) & --\\
V                & 1800 (2)  &  1600 (3)  & --          & 1000 (2)\\
I                 & 1800 (4)  &  1470 (4)  & 800 sec (1) &  460 (4)\\
\noalign{\smallskip} \hline
\end{tabular}
\end{table} 
\clearpage
\textheight 25cm
\begin{deluxetable}{l c c c c c c c c }
\centering
\tabletypesize{\footnotesize}
\tablecaption{Photometric data for NGC~4472. The coordinates are given for J2000.  \label{photo4472}}
\startdata
  ID & RA [h:m:s]& DEC [d:m:s]& V [mag]& $\sigma$V [mag]& I [mag] & $\sigma$I [mag] & K [mag] & $\sigma$K [mag] \\
\hline
  1  &12:29:43.0 &  7:59:10.9 & 22.350 & 0.030 &  21.080 &  0.030 &  19.074 &  0.095    \\
  2  &12:29:43.1 &  8:00:40.2 & 21.110 & 0.010 &  20.100 &  0.010 &  18.575 &  0.060    \\
  3  &12:29:43.3 &  8:00:22.1 & 21.350 & 0.020 &  20.360 &  0.020 &  18.911 &  0.082    \\
  4  &12:29:43.5 &  8:00: 7.3 & 22.110 & 0.030 &  20.920 &  0.030 &  19.280 &  0.114    \\
  5  &12:29:43.5 &  7:59:49.7 & 22.240 & 0.030 &  20.930 &  0.030 &  18.971 &  0.086    \\
  6  &12:29:43.6 &  8:00:53.4 & 21.860 & 0.020 &  20.690 &  0.020 &  19.144 &  0.101    \\
  7  &12:29:44.5 &  7:59:19.9 & 22.230 & 0.020 &  21.240 &  0.020 &  19.549 &  0.146    \\
  8  &12:29:44.6 &  7:59:54.6 & 22.380 & 0.050 &  21.160 &  0.040 &  19.238 &  0.110    \\
  9  &12:29:44.6 &  7:59:35.9 & 21.180 & 0.020 &  20.150 &  0.020 &  19.038 &  0.091    \\
  10 &12:29:45.1 &  7:59:50.9 & 21.830 & 0.030 &  20.510 &  0.030 &  18.694 &  0.067    \\
  11 &12:29:45.4 &  7:59:37.0 & 21.390 & 0.020 &  20.130 &  0.020 &  18.204 &  0.043    \\
  12 &12:29:46.0 &  7:59:17.4 & 21.530 & 0.010 &  20.360 &  0.010 &  18.540 &  0.058    \\
  13 &12:29:46.3 &  7:59:49.3 & 21.380 & 0.010 &  20.240 &  0.010 &  18.658 &  0.065    \\
  14 &12:29:46.5 &  8:00:34.5 & 21.730 & 0.030 &  20.450 &  0.020 &  18.522 &  0.057    \\
  15 &12:29:46.6 &  7:59:41.0 & 21.220 & 0.020 &  20.070 &  0.030 &  16.193 &  0.008    \\
  16 &12:29:46.6 &  8:00:33.8 & 21.370 & 0.020 &  20.060 &  0.020 &  18.231 &  0.044    \\
  17 &12:29:46.8 &  8:00:25.1 & 21.760 & 0.040 &  20.500 &  0.040 &  19.029 &  0.091    \\
  18 &12:29:47.3 &  7:59:45.7 & 21.210 & 0.010 &  19.940 &  0.010 &  18.129 &  0.040    \\
  19 &12:29:47.3 &  8:00: 0.0 & 22.100 & 0.030 &  20.980 &  0.040 &  18.857 &  0.078    \\
  20 &12:29:47.4 &  7:59:12.0 & 22.130 & 0.010 &  20.900 &  0.010 &  19.416 &  0.129    \\
  21 &12:29:47.4 &  8:00: 4.5 & 23.450 & 0.100 &  22.210 &  0.090 &  19.091 &  0.096    \\
  22 &12:29:47.5 &  7:59:36.1 & 22.180 & 0.030 &  20.990 &  0.030 &  19.429 &  0.131    \\
  23 &12:29:47.7 &  8:00:24.7 & 21.810 & 0.020 &  20.890 &  0.020 &  19.394 &  0.127    \\
  24 &12:29:47.7 &  7:59:53.6 & 22.420 & 0.020 &  21.310 &  0.030 &  19.478 &  0.137    \\
  25 &12:29:47.8 &  7:59:26.3 & 22.630 & 0.030 &  21.360 &  0.030 &  19.349 &  0.121    \\
  26 &12:29:47.9 &  7:59:19.6 & 21.330 & 0.010 &  20.130 &  0.010 &  18.411 &  0.052    \\
  27 &12:29:48.1 &  8:00:21.2 & 22.220 & 0.020 &  20.900 &  0.020 &  19.046 &  0.092    \\
  28 &12:29:49.1 &  7:59:53.6 & 21.680 & 0.010 &  20.500 &  0.010 &  19.092 &  0.096    \\
  29 &12:29:49.1 &  8:00:35.0 & 22.510 & 0.020 &  21.200 &  0.020 &  19.114 &  0.098    \\
  30 &12:29:49.2 &  8:00:56.2 & 21.350 & 0.010 &  20.060 &  0.010 &  17.947 &  0.034    \\
  31 &12:29:49.5 &  7:59:35.1 & 21.840 & 0.010 &  20.600 &  0.010 &  18.782 &  0.072    \\
  32 &12:29:49.6 &  7:59:40.7 & 21.970 & 0.010 &  20.960 &  0.010 &  19.485 &  0.138    \\
  33 &12:29;49.8 &  7:59:16.2 & 21.870 & 0.010 &  20.700 &  0.010 &  19.001 &  0.088    \\
  34 &12:29:50.0 &  7:59:36.1 & 23.100 & 0.030 &  21.750 &  0.030 &  19.517 &  0.142    \\
  35 &12:29:50.0 &  7:59:43.7 & 21.970 & 0.010 &  20.730 &  0.010 &  19.204 &  0.106    \\
  36 &12:29:50.1 &  7:59:44.2 & 21.410 & 0.010 &  20.280 &  0.010 &  18.928 &  0.083    \\
  37 &12:29:50.4 &  8:00:14.2 & 21.260 & 0.010 &  19.960 &  0.010 &  17.761 &  0.029    \\
\enddata
\end{deluxetable}
\textheight 20cm
\begin{deluxetable}{l c c c c c c c c }
\centering
\tabletypesize{\footnotesize}
\tablecaption{Photometric data for NGC~4594. The coordinates are given for J2000. \label{photo4594}}
\startdata
  ID & RA [h:m:s]& DEC [d:m:s]& V [mag]& $\sigma$V [mag]& I [mag] & $\sigma$I [mag] & K [mag] & $\sigma$K [mag]\\
\hline
1 & 12:40: 0.3 & -11:37:21.6 &  19.401  &  0.008 &  18.138  &  0.006 &  16.091 &   0.010 \\
2 & 12:39:59.3 & -11:37:12.2 &  21.250  &  0.031 &  20.200  &  0.028 &  18.906 &   0.097 \\
3 & 12:39:58.7 & -11:37:17.8 &  20.417  &  0.018 &  19.321  &  0.014 &  17.440 &   0.031 \\
4 & 12:39:58.6 & -11:37:26.0 &  20.412  &  0.014 &  19.022  &  0.011 &  17.311 &   0.044 \\
5 & 12:39:58.9 & -11:37:43.1 &  20.306  &  0.013 &  19.024  &  0.008 &  17.116 &   0.020 \\
6 & 12:40: 2.1 & -11:37:21.7 &  20.997  &  0.029 &  19.686  &  0.023 &  17.744 &   0.048 \\
7 & 12:40: 2.1 & -11:37: 8.8 &  20.344  &  0.023 &  19.114  &  0.018 &  17.377 &   0.031 \\
8 & 12:40: 2.5 & -11:37: 7.4 &  21.193  &  0.019 &  19.941  &  0.016 &  17.911 &   0.039 \\
9 & 12:40: 0.9 & -11:37: 6.6 &  19.836  &  0.010 &  18.650  &  0.009 &  16.913 &   0.019 \\
10 & 12:40: 1.0 & -11:36:54.2 &  20.443 &   0.011 &  19.428 &   0.009 &  18.070 &   0.060 \\
11 & 12:40: 2.6 & -11:36:40.9 &  21.182 &   0.016 &  19.977 &   0.011 &  18.300 &   0.058 \\
12 & 12:40: 0.7 & -11:36:49.6 &  21.053 &   0.018 &  20.020 &   0.015 &  18.686 &   0.121 \\
13 & 12:40: 0.6 & -11:36:45.4 &  20.883 &   0.015 &  19.738 &   0.010 &  18.192 &   0.066 \\
14 & 12:40: 2.6 & -11:36:27.6 &  21.155 &   0.016 &  20.147 &   0.012 &  19.192 &   0.128 \\
15 & 12:39:59.3 & -11:36:30.4 &  20.650 &   0.012 &  19.445 &   0.007 &  17.784 &   0.047 \\
16 & 12:40: 2.2 & -11:37:59.8 &  19.658 &   0.007 &  18.386 &   0.005 &  16.146 &   0.008 \\
17 & 12:40: 3.1 & -11:37:19.8 &  22.602 &   0.095 &  21.179 &   0.056 &  18.734 &   0.073 \\
18 & 12:39:58.7 & -11:37:47.4 &  20.356 &   0.012 &  19.356 &   0.009 &  17.969 &   0.063 \\
19 & 12:39:59.8 & -11:37:55.7 &  22.107 &   0.041 &  20.857 &   0.027 &  18.610 &   0.128 \\
20 & 12:40: 0.2 & -11:37:53.3 &  20.052 &   0.009 &  19.049 &   0.007 &  17.620 &   0.038 \\
21 & 12:40: 1.0 & -11:37:55.5 &  20.176 &   0.010 &  18.932 &   0.006 &  16.966 &   0.018 \\
22 & 12:39:58.2 & -11:38:18.9 &  21.237 &   0.016 &  20.142 &   0.012 &  18.456 &   0.070 \\
23 & 12:39:58.5 & -11:38:20.1 &  21.789 &   0.026 &  20.613 &   0.016 &  19.046 &   0.147 \\
24 & 12:40: 1.8 & -11:38: 2.9 &  21.888 &   0.029 &  20.598 &   0.018 &  18.719 &   0.102 \\
25 & 12:39:59.2 & -11:38:20.0 &  20.969 &   0.015 &  19.798 &   0.009 &  17.960 &   0.048 \\
26 & 12:40: 1.7 & -11:38:22.1 &  21.399 &   0.019 &  20.420 &   0.014 &  19.167 &   0.142 \\
\enddata
\end{deluxetable}

\begin{deluxetable}{l c c c c c c c c }
\centering
\tabletypesize{\footnotesize}
\tablecaption{Photometric data for NGC~3585. The coordinates are given for J2000.  \label{photo3585}}
\startdata
  ID & RA [h:m:s]& DEC [d:m:s]& B [mag]& $\sigma$B [mag]& I [mag] & $\sigma$I [mag] & K [mag] & $\sigma$K [mag]\\
\hline
1 & 11:13:17.3&  -26:46: 9.9&   22.990&    0.019&   20.813&    0.010&   18.975&    0.052\\
2 & 11:13:19.1&  -26:46: 4.6&   23.043&    0.020&   20.859&    0.010&   18.999&    0.053\\
3 & 11:13:15.9&  -26:45:56.6&   22.414&    0.012&   20.778&    0.009&   18.301&    0.029\\
4 & 11:13:20.2&  -26:45:50.7&   18.567&    0.001&   16.756&    0.001&   14.935&    0.002\\
5 & 11:13:15.7&  -26:45:53.5&   22.977&    0.019&   20.795&    0.009&   17.976&    0.021\\
6 & 11:13:13.8&  -26:45:51.1&   22.700&    0.015&   20.549&    0.008&   18.604&    0.037\\
7 & 11:13:18.9&  -26:45:42.2&   23.373&    0.026&   21.241&    0.014&   19.275&    0.068\\
8 & 11:13:20.2&  -26:45:38.3&   21.969&    0.009&   19.836&    0.004&   17.810&    0.018\\
9 & 11:13:16.3&  -26:45:37.7&   22.529&    0.013&   20.375&    0.007&   18.609&    0.037\\
10 & 11:13:15.7&  -26:45:29.1&   24.086&    0.049&   21.373&    0.016&   19.199&    0.064\\
11 & 11:13:15.9&  -26:45:23.5&   22.097&    0.010&   19.864&    0.004&   17.127&    0.010\\
12 & 11:13:14.7&  -26:45:22.6&   23.055&    0.020&   20.855&    0.010&   19.105&    0.059\\
13 & 11:13:14.3&  -26:45:18.3&   23.848&    0.039&   21.687&    0.020&   19.604&    0.092\\
14 & 11:13:16.4&  -26:45: 6.1&   22.789&    0.016&   20.520&    0.007&   19.236&    0.066\\
15 & 11:13:19.1&  -26:45: 4.9&   23.210&    0.023&   21.117&    0.012&   19.478&    0.082\\
16 & 11:13:14.5&  -26:45: 2.3&   23.485&    0.028&   21.366&    0.015&   20.084&    0.142\\
17 & 11:13:16.1&  -26:45: 1.1&   23.060&    0.022&   19.419&    0.003&   17.653&    0.016\\
18 & 11:13:16.8&  -26:44:57.7&   20.475&    0.003&   18.212&    0.001&   16.686&    0.007\\
19 & 11:13:13.4&  -26:44:58.8&   23.021&    0.019&   21.258&    0.014&   19.881&    0.111\\
20 & 11:13:16.9&  -26:44:52.0&   22.209&    0.011&   19.055&    0.002&   17.191&    0.011\\
21 & 11:13:18.8&  -26:44:47.1&   22.663&    0.014&   20.698&    0.009&   19.082&    0.057\\
22 & 11:13:18.9&  -26:44:47.1&   23.753&    0.039&   20.271&    0.006&   18.454&    0.033\\
23 & 11:13:16.7&  -26:44:41.7&   23.199&    0.022&   21.282&    0.014&   19.838&    0.114\\
24 & 11:13:15.7&  -26:44:31.9&   22.751&    0.016&   20.570&    0.008&   18.894&    0.048\\
25 & 11:13:19.8&  -26:44:26.5&   22.029&    0.009&   19.803&    0.004&   17.610&    0.015\\
26 & 11:13:15.1&  -26:44:28.7&   24.754&    0.086&   22.582&    0.046&   19.786&    0.108\\

\enddata
\end{deluxetable}

\begin{deluxetable}{l c c c c c c c c}
\centering
\tabletypesize{\footnotesize}
\tablecaption{Photometric data for NGC~5813. The coordinates are given for J2000. \label{photo5813}}
\startdata
\hline
  ID & RA [h:m:s]& DEC [d:m:s]& V [mag]& $\sigma$V [mag]& I [mag] & $\sigma$I [mag] & K [mag] & $\sigma$K [mag]\\
\hline
1& 15: 1:10.1 & 1:42: 6.4 &  21.982  & 0.020 &  20.908  & 0.031 &  19.461  & 0.134 \\
2& 15: 1:12.1 & 1:42:13.7 &  22.030  & 0.021 &  20.942  & 0.033 &  19.386  & 0.103 \\
3& 15: 1:12.2 & 1:41:59.5 &  21.475  & 0.015 &  19.740  & 0.013 &  18.087  & 0.030 \\
4& 15: 1:10.3 & 1:42:42.2 &  24.296  & 0.093 &  22.058  & 0.063 &  18.837  & 0.047 \\
5& 15: 1: 9.6 & 1:42:43.2 &  24.686  & 0.137 &  22.086  & 0.058 &  18.744  & 0.044 \\
6& 15: 1: 8.6 & 1:42:59.0 &  22.994  & 0.029 &  21.764  & 0.040 &  19.749  & 0.142 \\
7& 15: 1: 8.8 & 1:42:52.8 &  21.949  & 0.016 &  21.022  & 0.025 &  19.694  & 0.141 \\
8& 15: 1: 9.8 & 1:42:27.0 &  20.913  & 0.009 &  20.036  & 0.014 &  19.162  & 0.084 \\
9& 15: 1: 8.7 & 1:42:45.4 &  22.366  & 0.020 &  21.342  & 0.033 &  19.680  & 0.135 \\
10& 15: 1: 8.6 & 1:42:36.4 &  22.049  & 0.016 &  20.963  & 0.023 &  19.411  & 0.090 \\
11& 15: 1: 8.7 & 1:42: 9.2 &  21.722  & 0.014 &  20.753  & 0.023 &  19.015  & 0.055 \\
12& 15: 1:11.4 & 1:42:41.9 &  23.600  & 0.066 &  22.647  & 0.098 &  19.576  & 0.100 \\
13& 15: 1:11.4 & 1:42:42.2 &  22.320  & 0.021 &  21.298  & 0.031 &  19.576  & 0.100 \\
14& 15: 1:12.4 & 1:42:54.7 &  21.681  & 0.013 &  20.654  & 0.018 &  19.036  & 0.065 \\
15& 15: 1:11.4 & 1:42:55.8 &  23.654  & 0.054 &  21.574  & 0.035 &  18.433  & 0.035 \\
16& 15: 1:12.1 & 1:43: 3.4 &  22.174  & 0.016 &  21.095  & 0.024 &  19.623  & 0.128 \\
17& 15: 1:12.4 & 1:42:22.5 &  21.404  & 0.012 &  20.447  & 0.020 &  18.911  & 0.066 \\
18& 15: 1:12.1 & 1:42:35.3 &  21.622  & 0.014 &  20.620  & 0.020 &  19.233  & 0.078 \\
19& 15: 1:13.6 & 1:42: 8.2 &  21.703  & 0.015 &  20.673  & 0.021 &  19.233  & 0.084 \\
20& 15: 1:13.1 & 1:42:31.2 &  22.914  & 0.036 &  20.757  & 0.026 &  17.160  & 0.009 \\
21& 15: 1:14.4 & 1:42:10.4 &  21.167  & 0.010 &  20.092  & 0.015 &  18.387  & 0.034 \\
22& 15: 1:13.5 & 1:42:40.5 &  21.939  & 0.017 &  20.912  & 0.025 &  19.304  & 0.082 \\

\enddata
\end{deluxetable}

\end{document}